\documentclass[printer]{aa}

\usepackage{graphicx}
\usepackage[varg]{txfonts}
\usepackage{natbib}
\usepackage{orcidlink}

\usepackage{hyperref}
\hypersetup{
    colorlinks=true,
    linkcolor=cyan,
    urlcolor=blue,
    citecolor=black
    }
    
\usepackage[normalem]{ulem}

\newcommand{\solrad}{$\mathrm{R}_\odot$} 
\newcommand{\kms}{$\mathrm{km}\,\mathrm{s}^{-1}$}
\begin{document}

   \title{Estimating uncertainties in the back-mapping of the fast solar wind}
   
   \titlerunning{Uncertainties in the back-mapping of solar wind}
   \authorrunning{Koukras et al.}

   \author{A. Koukras
          \inst{1,}\inst{2} \orcidlink{0000-0002-6232-5527}
          \and L. Dolla\inst{2}\orcidlink{0000-0001-6353-5887} 
          \and R. Keppens\inst{1}\orcidlink{0000-0003-3544-2733}
          }

   \institute{Centre for mathematical Plasma-Astrophysics, Department of Mathematics, KU Leuven, Celestijnenlaan 200B, 3001 Leuven, Belgium\\
             \email{alexandros.koukras@kuleuven.be}\\
             \email{rony.keppens@kuleuven.be}
         \and
             Solar-Terrestrial Centre of Excellence, Royal Observatory of Belgium, Avenue Circulaire 3, B-1180 Brussels, Belgium\\
             \email{laurent.dolla@oma.be}
             }

   \date{Received ..; accepted ....}

  \abstract
    {Although the most likely source regions of fast solar wind relate to coronal holes, the exact acceleration mechanism that drives the fast solar wind is still not fully understood. An important approach that can improve our understanding is the combination of remote sensing and in situ measurements, which is often referred to as linkage analysis. This linkage tries to identify the source location of the in situ solar wind with a process called back-mapping. Typically, back-mapping is a combination of ballistic mapping, where the solar wind draws the magnetic field into the Parker Spiral at larger radial distances and magnetic mapping, where the solar wind follows the magnetic field line topology from the solar surface to a point in the corona where the solar wind starts to expand radially. }
    {By examining the different model ingredients that can affect the derived back-mapped position, we aim to provide a more precise estimate of the source location and a measure of confidence in the mapping procedure. This can be used to improve the connection of remote sensing with in situ measurements.}
    {For the ballistic mapping we created velocity profiles based on Parker wind approximations. These profiles are constrained by observations of the fast solar wind close to the Sun and are used to examine the mapping uncertainty. 
    The coronal magnetic field topology from the solar surface up to an outer surface (source surface; SS) radius $R_{SS}$ is modeled with a potential field source surface extrapolation (PFSS). PFSS takes as input a photospheric synoptic magnetogram and a value for the source surface radius, which is defined as the boundary after which the magnetic field becomes radial. The sensitivity of the extrapolated field is examined by adding reasonable noise to the input magnetogram and performing a Monte Carlo simulation, where for multiple noise realizations we calculate the source position of the solar wind. Next, the effect of free parameters, like the height of the source surface, is examined and statistical estimates are derived. We used Gaussian Mixture clustering to group the back-mapped points, due to different sources of uncertainty, and provide a confidence area for the source location of the solar wind. 
    Furthermore, we computed a number of metrics to evaluate the back-mapping results and assessed their statistical significance by examining 3 high speed stream events. Lastly, we explored the effect of corotation, close to the Sun, on the source region of the solar wind. }
    {For back-mapping with a PFSS corona and ballistic solar wind, our results show that the height of the source surface produces the largest uncertainty in the source region of the fast solar wind, followed by the choice of the velocity profile and the noise in the input magnetogram. Additionally, we display the ability to derive a confidence area on the solar surface that represents the potential source region of the in-situ measured fast solar wind.}
    {}

    \keywords{The Sun -- Sun: magnetic fields -- solar wind
                }
                 

  \maketitle

\section{Introduction}\label{Introduction}
The solar wind is a continuous magnetized plasma flow emanating from the Sun, turning supersonic and super-Alfv\'enic at relatively close distances.
It is the extension of the solar atmosphere which expands into the interplanetary space. 
It was first theorized based on the observed effects on cometary tails \citep{Eddington1910,Biermann1957} and on the interactions with the Earth's magnetic field \citep{Birkeland1914}, as corpuscular radiation. But it was \citet{Parker1958}, who by using the first observational evidence of a hot solar corona \citep{Edlen1943,Alfven1947,Chapman1957} deduced that such hot solar atmosphere can not maintain hydrostatic equilibrium and the plasma flow from the Sun must expand and turn supersonic. He coined the term `solar wind' and soon afterwards in situ data by the Mariner II spacecraft confirmed his prediction \citep{Neugebauer1962}.

From this point onward there have been consistent measurements of the solar wind, either in situ with spacecrafts like Ulysses \citep{Bame1992}, ACE \citep{Stone1998} and WIND \citep{Acuna1995} or remotely with UVCS \citep{Kohl1995} and LASCO \citep{Brueckner1995}, which provided a detailed view of the solar wind properties. These revealed a categorization of the solar wind based on its speed in slow ($<500$~\kms) and fast wind ($>500$~\kms) \citep{Schwenn1990,Geiss1995,Zurbuchen2007}. There has been extensive discussion whether other solar wind properties could provide a more reliable categorization \citep{Neugebauer2016,Camporeale2017} but a generic classification remains based on typical fast-wind properties and typical slow-wind properties, encompassing additional plasma parameters such as ion composition, Alfv\'enicity and first ionization potential (FIP) bias \citep{Verscharen2019}. 

From the first Ulysses data it was evident that the fast solar wind originated from dark regions on the Sun near the poles, called coronal holes \citep{Krieger1973}, and the slow wind from within or near the partially closed streamer belt, at the solar equator \citep{Schrijver2009book,Verscharen2019}. 
Subsequent observations enhanced this view and showed that at solar maximum the fast and slow solar wind can emerge from everywhere in the corona in neighboring patches \citep{Verscharen2019}. 
For a more in depth view on the nature of the solar wind see reviews by \citet{Antiochos2012,Cranmer2017,Verscharen2019} and references therein.

\begin{figure}
    \centering
    \includegraphics[scale=0.3]{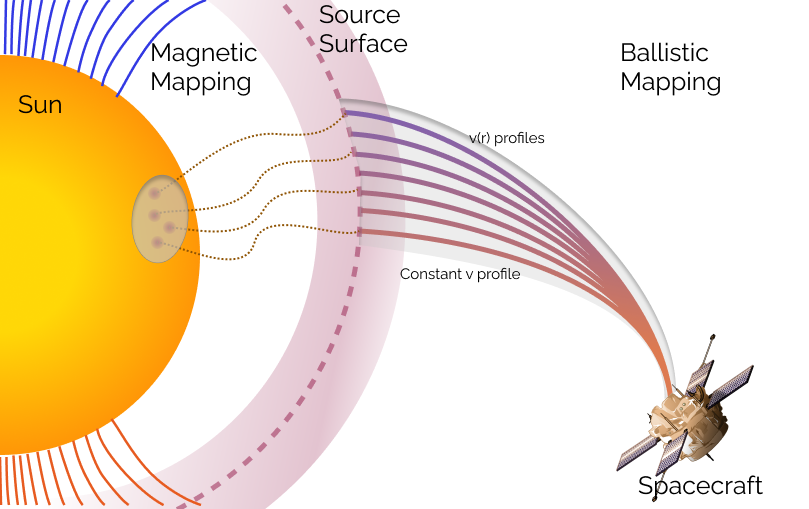} 
    \caption{Representation of the solar wind back-mapping process and its uncertainties assessed in this paper. In the zone of ballistic mapping (above the source surface), the lines that connect the spacecraft to the source surface represent different trajectories for different velocity profiles of the solar wind; the shaded area around them represents the uncertainty associated with them. The shaded area around the dashed circular line (which shows the source surface) represents the uncertainty in its height. In the magnetic mapping zone (below the source surface) the magnetic field lines are followed from the source surface down to the solar surface (dotted lines). The shaded area around the back-mapped positions on the solar surface represents the uncertainty in the source region of the wind observed at the spacecraft.}
    \label{fig:sw_cartoon}
\end{figure}

In order to derive the source location on the solar surface, of an in situ solar wind measurement, one of the most common methods is based on a process called solar wind back-mapping \citep{Neugebauer1998,Peleikis2017,Badman2020}. 
Back-mapping consists of two main parts and is often referred to as two-step ballistic mapping. The first is the ballistic mapping, where the solar wind is traced from the in situ point to a point in the outer corona, which is called source surface (SS) and it is the boundary above which the modeled coronal magnetic field becomes radial. 
The second is the magnetic mapping which continues the tracing from the source surface down to the photosphere. A graphical example of the back-mapping process can be seen in Fig. \ref{fig:sw_cartoon}.
This figure was inspired from Fig. 1 in \cite{Peleikis2017} and together with the back-mapping process it displays the uncertainties in different components of the framework, which are indicated with the shaded areas.

As \citet{Parker1958} showed the solar wind becomes nearly radial after a critical point close to the Sun, and pulls the magnetic field into a Parker spiral. The ballistic mapping that has been used to date usually assumes that the in situ measured speed remains constant along the entire Parker spiral. Based on this assumption, if we have an in situ speed $v_{sc}$, measured from a spacecraft at radial distance from the Sun $r_{sc}$ and the radial extend of the source surface $r_{ss}$, we can compute the time it takes the solar wind to reach the source surface ($\Delta t$, the so-called back-mapping time) and the corresponding displacement in Carrington longitude of its footpoint $\phi_{ss}$ on the source surface
\begin{eqnarray}
    \Delta t &  = & \frac{r_{sc} - r_{ss}}{v_{sc}} \,,\\
    \Delta \phi & = & \Omega \ \Delta t \,,\\
    \phi_{ss} & = & \phi_{sc} + \Delta \ \phi \,,
\end{eqnarray}
where $\Omega$ is the solar sidereal rotation rate and $\phi_{sc}$ the Carrington longitude of the spacecraft. In this formulation, the spacecraft is orbiting the Sun and its position is given in the Carrington coordinate system, which is a system that co-rotates with the Sun. Lastly, the ballistic mapping does not produce any latitudinal displacement $\theta_{sc} = \theta_{ss}$, also the azimuthal component of the solar wind velocity vector is considered negligible and it is not taken into consideration (see Sect.~\ref{Discussion}).
The aforementioned assumption of constant speed propagation is based mainly on the work of \citet{Nolte1973a}, who showed that such an approximation derives a fairly accurate estimation of the source region.
In this paper, we will explore more realistic means for ballistic back-mapping, where we account for the solar wind speed variation as a function of distance, and do this based on observationally constrained wind profiles.

For the magnetic mapping a derivation of the coronal magnetic field topology is necessary, this is achieved by global models, such as the Potential Field Source Surface (PFSS) model \citep{Altschuler1969,Schatten1969}, which extrapolates the photospheric magnetic field in the corona up until a certain height which represents the source surface (SS) (the point after which the magnetic field is assumed to become radial). 
In this domain, solar wind can be connected from the source surface down to the solar surface by tracing the magnetic field lines. 
This part of the back-mapping may be influenced by errors in the input data and the variation of the free parameters in the extrapolation model. This will be examined here as well.
For this study we used synoptic magnetograms from the Global Oscillation Network Group (GONG;\citealt{Harvey1996}) as they are widely used as input for magnetic extrapolation models and space weather modeling in general \citep{Plowman2020a}. 
Additionally, different magnetogram sources do not have strong deviations compared to GONG in back-mapping frameworks, and results based on GONG magnetograms may be preferred \citep{Badman2020}. 

Despite the wide use of the back-mapping method there has been limited work on estimating the uncertainties in its implementation. 
\citet{Peleikis2017} added noise with a uniform distribution (magnitude $\pm 10^{\circ}$) to simulate the error in the longitudinal displacement of the solar wind footpoint on the source surface, based on the $10^\circ$ error estimate of \citet{Nolte1973a}. Additionally, they added noise to each pixel in the input magnetograms (MDI), with a value of $\pm 0.1 \times I_p$, with $I_p$ the magnetic field strength at each pixel, to estimate an error from the PFSS extrapolation.  
\citet{Badman2020} used different sources of input magnetograms and heights of the source surface, to investigate the effect they have on reproducing the radial magnetic field measured from Parker Solar Probe (PSP).
And more recently \citet{MacNeil2022} investigate the error in the ballistic mapping proposed by \cite{Nolte1973a}, by comparing the simple ballistic approximation with models that include radial and azimuthal variations. This analysis was focused on the slow solar wind as the evaluation method was the comparison between crossings of the heliospheric current sheet (HCS) at Earth and at 2.5 \solrad{}. 

The importance of a detailed estimation of all the uncertainties involved in the back-mapping process becomes even more evident in the context of `linkage' analysis, where in situ measurements are connected to remote sensing measurements.
In this study we quantify the uncertainty in the source region location based on different components of the back-mapping framework and derive some quantitative precision estimates where possible.
The structure of the paper is as follows. 
In Sect. \ref{Methods_and_Analysis} we present our methodology, which  includes the description of the data set (Sect. \ref{Dataset}), the ballistic mapping with custom velocity profiles (Sect. \ref{Ballistic_Mapping}), the magnetic mapping (Sect. \ref{Magnetic_mapping}) and the different sources of uncertainty (Sect. \ref{Uncertainty_sources}). In Sect. \ref{Discussion} we discuss the results of our analysis and lastly in Sect. \ref{Conclusions} we provide our conclusions. 

\section{Methods and Analysis}\label{Methods_and_Analysis}

\subsection{Data set}\label{Dataset}

\begin{figure*}[h]
    \centering
    \includegraphics[scale=0.45]{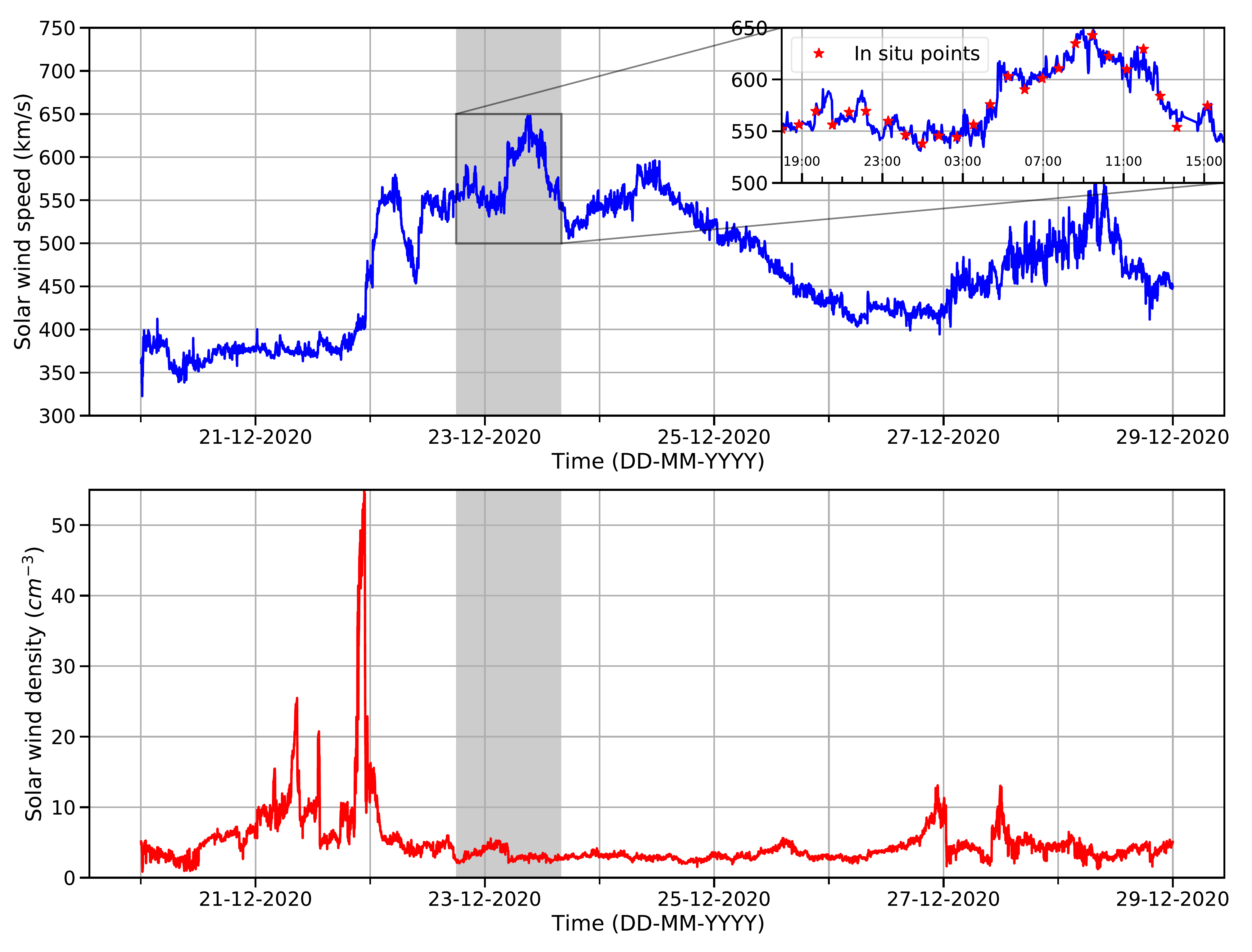}
    \caption{(Top) In situ solar wind speed for a case study high speed stream observed by WIND (Event 1). (Bottom) In situ solar wind density. The shaded area in both plots represents the high speed interval that was used during the analysis (Event 1). (Inset) The zoomed-in view of the selected high speed interval. The red star markers represent the in situ speed measurements that were back-mapped to the solar surface and examined in more detail.}
    \label{fig:sw_v-d_ev4}
\end{figure*}

In order to test our back-mapping framework, we looked at a number of events of fast solar wind originating from low latitude coronal holes. To demonstrate our methodology and analysis we will present in more detail a single case study of a high speed stream event that was observed on 21-25/12/2020. 
The data for this event are taken (with the use of HelioPy \citep{Stansby2020heliopy}) from the Solar Wind Experiment (SWE) on board the WIND spacecraft, which is located in a halo orbit at the Lagrangian point L1 and can been seen in Fig. \ref{fig:sw_v-d_ev4}. 
The top plot displays the solar wind speed and the bottom plot the solar wind density. The shaded area represents the high speed stream interval we studied. In this interval an average wind speed close to 600 \kms and a max speed of 650 \kms are observed. The inset plot displays a zoomed-in view of the selected interval. The red stars in this plot represent the in situ solar wind speed measurements that we examined in more detail. Since SWE provides measurements of the solar wind speed every 92 s we downsampled the selected interval in order to have a more manageable data set for our analysis. We selected a downsampling step of 30 measurements which is equivalent to a cadence of 46 min and a total number of 25 speed measurements for this event. We will refer to these speed measurements as "in-situ points" from now.

This particular event was selected because during this interval there were data available from a similar back-mapping framework, called Magnetic Connectivity Tool (MCT) \citep{Rouillard2017}, which is one of the components we use to evaluate our framework. Additionally, the wind speed for this event is not that large compared to the fastest solar wind streams that are observed, which provides the opportunity to examine the performance of our framework in the limit of what is considered fast solar wind and provide an upper bound on the possible uncertainty in the back-mapped region. Lastly, solar wind speeds in this range are correlated to the smallest fast wind speeds for which observational constraints from Doppler dimmings are available, these observations are discussed in Sect. \ref{velocity_profiles}.

In order to make sure we have a `pure' fast solar wind event, we avoided any compression regions as can be seen in the bottom plot of Fig. \ref{fig:sw_v-d_ev4} and we cross-checked with a near-Earth Interplanetary Coronal Mass Ejections (ICME) catalog\footnote{http://www.srl.caltech.edu/ACE/ASC/DATA/level3/icmetable2.htm} 
that there were no ICMEs during this high speed stream.
We will examine in much detail this fast wind event to present our methodology, mainly through the analysis of a single in situ point of this event. A statistical study of additional fast wind events will be then made, comprising, each, several in situ points. The details of these events can be seen in the Table \ref{table:1}. 

\begin{table}[h]
\caption{Fast solar wind events}              
\label{table:1}      
\centering                                      
\begin{tabular}{c c c c}          
\hline\hline                        
Event & Start & End & In-situ points \\    
\hline                                   
    1 & 22/12/2020 17:59 & 23/12/2020 15:10 & 25 \\      
    2 & 13/04/2012 04:07 & 13/04/2012 23:13 & 24 \\
    3 & 04/08/2017 13:19 & 05/08/2017 10:11 & 26 \\
\hline                                             
\end{tabular}
\end{table}

\subsection{Ballistic mapping} \label{Ballistic_Mapping}
The first implementation of ballistic mapping appears in \citet{Snyder_Neugebauer1966}, where based on the assumption of a constant solar wind speed and radial velocity they attempt to locate, following `ideal' Parker spirals, the sources of high speed streams observed by Mariner II \citep{Nolte1973a}.
Theoretical work in the 70s \citep{Sakurai1971,Matsuda1972} demonstrated that the solar wind propagates radially and with constant speed beyond the critical point in the quasi-radial hypervelocity approximation (QRH). The QRH approximation is based on the assumption that the sonic and Alfv\'en Mach numbers are large, and that gravitational potential and azimuthal convection effects are negligible. This approximation has shown to be valid for radial distances larger than 30 \solrad.

Based on that, \citet{Nolte1973a} introduced the extrapolated quasi-radial hypervelocity approximation (EQRH), where the assumption of constant speed and radial velocity for the solar wind can be extended below 30 \solrad{}. This approximation is based on the canceling of two corrections (coronal corotation, interplanetary acceleration) with opposite effect on the derived source location.
While this model does not reflect a realistic velocity profile for the solar wind, it has been proven to be a good approximation for mapping the in situ wind to its source in the outer corona and it is considered the standard practice for ballistic mapping \citep{Neugebauer1998,Peleikis2017,Badman2020}. 

\citet{Nolte1973a} used the comparison with steady-state streamlines from theoretical solutions (both azimuthally dependent and independent) of the steady-state plasma equations to argue that the error, in the longitude of the source region, of his mapping method is of the order of $10^{\circ}$.
Given the underlying assumptions of this model (quiet-time coronal expansion, quasi-stationary wind) and the simplicity in the error estimation (the difference of two extreme values) we wanted to provide a more rigorous assessment of the radial velocity profile on back-mapping efforts. Therefore, we here investigate how an actual sampling of possible solar wind velocity profiles that match observational constraints can improve the uncertainty in the source region of the back-mapped solar wind.

There are a plethora of models that describe the solar wind propagation and the corresponding velocity profile. They include the original isothermal/polytropic models \citep{Parker1958}, the exospheric models, which approach the acceleration of the solar wind from a kinetic point of view \citep{Jockers1970,Maksimovic1997a,Zouganelis2004} and a family of more sophisticated models, that examine the effect of Alfv\'en waves on the acceleration of the solar wind \citep[e.g.][]{Hu1999a}. 
But in order to have more control on the velocity profiles shape, especially close to the Sun, we will here introduce
a family of hybrid velocity profiles for the solar wind propagation based on the \citet{Parker1958} approximations for small and large distances from the Sun. Also, the impact of realistic tangential speed profiles is examined later in Sect.~\ref{Discussion}.

Parker, taking the approximation that the corona is isothermal, derives an equation for
the solar wind, which has a specific sonic point. Since the Parker solution is governed by a transcendental equation it can be decomposed further into two analytic, closed-form approximations, one for large
distance from the Sun and one for small distances.

The main equations are the original transcendental equation for the isothermal (temperature $T_0$) wind profile
\begin{equation}
   (\frac{u^2}{a_0^2}) - \ln(\frac{u^2}{a_0^2})  =  4\ln\frac{r}{r_s} + 4 \frac{r}{r_s} -3 \,,\label{eq1}
   \end{equation}
which features the then constant sound speed and sonic point
\begin{eqnarray}
   a_0  & =  & (\frac{2kT_0}{m_p})^{1/2} \,,\label{eq2} \\
   r_s & = & \frac{GM}{2a_0^2} \label{eq3} \,.
   \end{eqnarray}
Instead, we will here make hybrid use of the large (Eq.~\ref{eq4}) and small distance (Eq.~\ref{eq5}) approximations, given by
   \begin{eqnarray}
   v & \approx & 2a_0 [\ln(r/r_s)]^{1/2} \,,\label{eq4} \\
   v & \approx & a_0 e^{3/2} e^{-2r_s/r} \,. \label{eq5}
\end{eqnarray}
Since our aim is to have a function that takes as input the in situ measured solar wind speed and produces different velocity profiles based on free adjustable parameters, we combine the small and large distance approximations to create a family of `hybrid' velocity profiles that differ from each other by the acceleration close to the Sun.
Therefore, we actually relax the `sonic point' meaning for $r_s$ and rather select the $r_s$ (in units of $\mathrm{R}_\odot$) as our free parameter, where we simply locally switch prescriptions between Eq.~\ref{eq4}-\ref{eq5}. For radial distances above this value, we use the large distance approximations and for radial distances below, the small distances approximation. The fact that this combination of approximate wind profiles is not an isothermal wind solution anymore (and has a local jump in its derivative) makes this a `hybrid' profile, which we further constrain by the measured speed value. To that end, we constrain the $a_0$ value based on the in situ measured speed ($v_{s/c}$) and the radial distance where this was measured ($r_{s/c}$), for both approximations. E.g., for a selected $r_s = 2.7 \  \mathrm{R}_\odot$ and an in-situ observed speed of 600 \kms, measured at L1 ($\approx$ 215  $\mathrm{R}_\odot$), we can find the appropriate $a_0$ value for the large distances approximation as
\begin{eqnarray}
    a_{0_{large}} & = & \frac{v_{s/c}}{2[\ln{(r_{s/c}/r_s)}]^{1/2}} \,, \label{eq6}\\
    v_{large}(r) & = & 2 a_{0_{large}} [\ln(r/r_s)]^{1/2} \,. \label{eq7}   
\end{eqnarray}
Similarly, we constrain $a_{0_{small}}$ and derive the small distances section of our `hybrid' profile. 

\begin{figure*}[h]
    \centering
    \includegraphics[width=\textwidth]{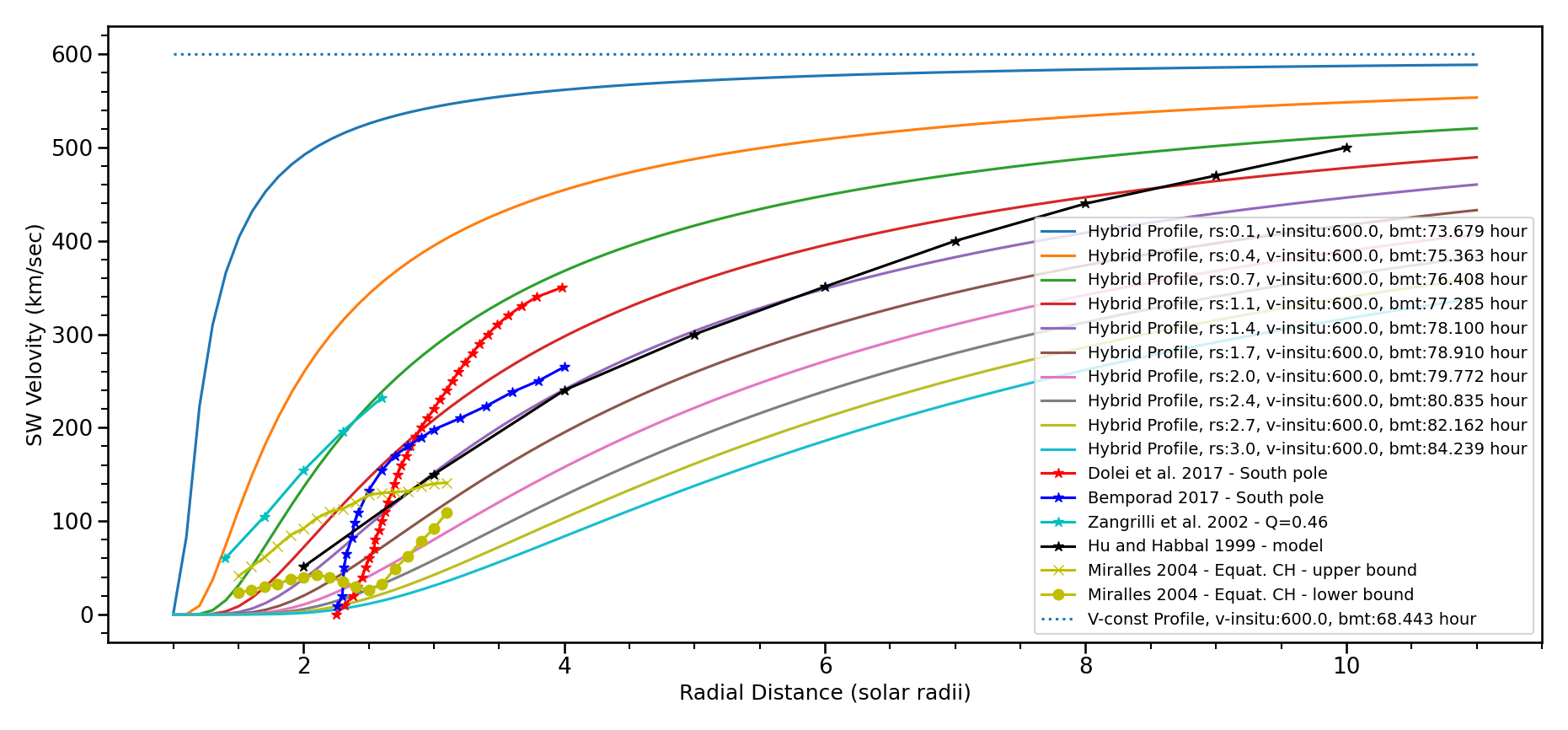}
    \caption{`Hybrid' solar wind velocity profiles for different values of the free parameters $r_s$, calculated for an in situ solar wind speed of 600 \kms \, (colored lines). Observations of solar wind outflow velocity close to the Sun, based on the Doppler dimming method are shown as lines with symbols. The velocity profile with constant speed is displayed by the horizontal dotted line. The profile output by the model of \citet{Hu1999} is also overploted, for comparison.}
    \label{fig:custom_profiles}
\end{figure*}

Hence, we create a solar wind velocity profile, combining the two Parker approximations, whose shape is determined by only one free parameter ($r_s$). In this formulation $r_s$ does not represent anymore physically the sonic point, but gives us a direct handle on the shape of the velocity profile. 
An example of these `hybrid' solar wind velocity profiles for different values of $r_s$ can be seen in Fig. \ref{fig:custom_profiles}, where all of these profiles actually match the local measured speed value at L1.

Summarizing, based on a single in situ solar wind speed measurement, we create a family of possible velocity profiles parameterized by $r_s$. For each profile, we can then determine the time it takes a solar wind parcel to travel from the source surface to the in situ measurement location (typically L1) by integrating its velocity profile, and we refer to this time as back-mapping time (from here on as bmt). 
It is evident that there is a lower limit on the back-mapping time (for every in situ solar wind speed), which is produced by assuming a constant speed all the way from the Sun. 
Since all the profiles we have in our family lie below the final constant speed, we need now to derive an upper limit on bmt, in order to define a confidence interval in the final source region location.
To accomplish that we look to constrain the space of the possible velocity profiles and by extension the space of possible $r_s$ values.

This effort of constraining the $r_s$ values is based on observations of the solar wind speed close to the Sun.
The acquisition of reliable measurements between 1-10 $\mathrm{R}_\odot$ is not a trivial task. There are 3 main remote sensing techniques, that are used for solar wind speed measurements in this region: the time delay of interplanetary radio scintillation, which is used to derive the speed of density irregularities typically in the region above 5 $\mathrm{R}_\odot$; the direct tracking of coronal features propagating in white light coronagraphic images; and lastly the Doppler dimming technique \citep{Withbroe1982}, which uses ultraviolet (UV) spectroscopic observations to derive solar wind outflow speeds \citep[and references within]{Bemporad2017}.

The UVCS (Ultra Violet Coronagraph Spectrometer) instrument \citep{Kohl1995} on board SOHO (Solar and Heliospheric Observatory) has performed systematic measurements of solar wind outflow speeds, using the Doppler dimming technique, for many years (different phases of the solar cycle) and it encompasses a multitude of solar latitudes. This plethora of measurements makes the results of the Doppler dimming method the more reliable observational constraint for our velocity profile space.
In this respect, observational solar wind velocity profiles that are associated with coronal holes \citep[fast solar wind origin; ][]{Zangrilli2002,Miralles2004,Bemporad2017,Dolei2018} can be used directly in our framework to define the largest $r_s$ value that produces the 'hybrid' velocity profile better matching the measured velocity profiles, which in turn will provide the upper limit in bmt. Furthermore, profiles from other models \citep[e.g. ][]{Hu1999} can also be included for comparison.


An example of how the observational constraints based on the Doppler dimming technique match with our family of hybrid profiles can be seen in Fig. \ref{fig:custom_profiles}.
In this figure, the colored lines represent a family of velocity profiles, computed for a single in situ speed value of 600 \kms. The different colors indicate different values of the free parameter $r_s$, as shown in the legend.
All of these profiles reach the same speed in situ at the spacecraft (at 215 \solrad{}), but here only a radial extent until 11 \solrad{} is displayed, for a better visual comparison with the remote-sensing measurements.
The horizontal dotted line represents the solar wind profile with constant speed (equal to the in situ measurement). As expected, this profile gives the shortest travel time for the solar wind.
Lastly, remote-sensing measurements, above coronal holes, of the solar wind outflow close to the Sun are presented with colored lines connecting different symbols. Specifically, lines that use the star symbol represent observational data from polar coronal holes.  
Here, we can notice how the $r_s$ values affect the `hybrid' profile shape and how the constant speed profile and the remote-sensing measurements bound the profile space. 
The profiles shown have $r_s$ varying from 0.1 to 3.0, leading to bmt values from 73 up to 84 hours, respectively and the shortest bmt of 68 hours is given by the constant profile.
\subsection{Magnetic Mapping}\label{Magnetic_mapping}

The second part of the back-mapping framework, as briefly mentioned in Sec.\ref{Introduction}, consists of the magnetic mapping of a solar wind parcel from the outer corona (source surface) %
down to the photosphere. The reason we back-map the solar wind through the tracing of the magnetic field lines in this region, is based on the local plasma conditions. In this region the plasma beta parameter $\beta = 2 \mu_0 {p}/{B^2}$ is usually smaller than unity, which means that magnetic forces dominate over pressure gradients and under the ideal magnetohydrodynamic (MHD) prescription we can assume that the outflow is constrained by the field topology.

In the Parker outflow model, the outward pressure of thermal plasma accelerates the solar wind and forces Sun's coronal magnetic field to open into the heliosphere.
This opening of the magnetic field can be approximated by PFSS models which impose that at a boundary surface, typically at a height of  2.5 \solrad, the magnetic field can have only a radial component.
At that height, the higher-order multipoles that describe the coronal field on scales associated with individual active regions have decreased so much that the field entering the heliosphere is dominated by the low-order dipole and quadrupole components of the global photospheric field. As a result, the pattern of the radial field at this height mostly consists of two large patches of opposite polarity, separated by an evolving, undulating, neutral line where the radial field itself vanishes.
This neutral line on the source surface extends into the heliosphere through the heliospheric current sheet.
Based on that, the selection of the appropriate source surface (SS) height is important, as it affects the overall topology. For example, for lower SS heights more coronal structures become open, which results in more open flux to extend in the interplanetary medium \citep{Lee2011}.

Specifically, the PFSS model represents the simplest case of a force-free model. Force-free models rely on a number of assumptions about the physical conditions in the corona. In ideal MHD, the momentum balance equation is 
\begin{equation}
    \rho (\frac{\partial \bold{v}}{\partial t} + \bold{v} \cdot \nabla \bold{v}) + \nabla p - \bold{J} \times \bold{B} + \rho \bold{g} = 0 \,.\label{eq8}
\end{equation}
Under the assumption that the corona is static on the photospheric time scale, which originates from the big difference in density between corona and photosphere ($n_{ph}/n_{c} \approx 10^8$), we can consider that the corona evolves as a series of quasi-static equilibria. 
Now, with the assumption that the plasma is static ($\partial/\partial t = 0$ and $\bold{v}=\bold{0}$) we arrive at the magnetohydrostatic momentum balance equation
\begin{equation}
    \nabla p - \bold{J} \times \bold{B} + \rho \bold{g} = 0 \,. \label{eq9}
\end{equation}
Next, if we consider a coronal domain with $\beta < 1$ where coronal structures change on length scales comparable to or shorter than the typical coronal scale height ($H = p/\rho g$), Eq. (\ref{eq9}) reduces to the force free equation 
\begin{eqnarray}
    \bold{J} \times \bold{B} & = & \bold{0} \,. \label{eq10} 
\end{eqnarray}
A simple solution for Eq. (\ref{eq10}) can be obtained by assuming that the current density vanishes. In this case we have a potential magnetic field. From Amp\`ere's law ($\nabla \times \bold{B} = \mu_0 \bold{J}$) we arrive at the current-free equation, coupled with the solenoidal condition for $\bold{B}$
\begin{eqnarray}
    \nabla \times \bold{B} & = & \bold{0} \,, \label{eq11}\\
    \nabla \cdot \bold{B} & = & 0 \,.\label{eq12}
\end{eqnarray}

The PFSS model solves for a magnetic field that satisfies Eq.~\ref{eq11}-\ref{eq12} in a spherical shell $1< r < R_{ss}$, between the photosphere (r=1\solrad{}) and an outer boundary at $r=R_{ss}$, which is the source surface. 
PFSS imposes two boundary conditions, the first is that at the source surface the magnetic field is strictly radial ($B_\phi = B_\theta = 0 $ on $r=R_{ss}$) 
and the second is that at the lower boundary the magnetic field equals the measured radial photospheric magnetic field ($B_r(\theta,\phi) = g(\theta,\phi)$ on $r=1$). 
Eq. (\ref{eq11}) means that we can express the magnetic field by the gradient of a scalar potential $\phi_B$, such that $\bold{B} = -\nabla \phi_B$. From the solenoidal condition (no-monopole) we have $\nabla \cdot \bold{B} = - \nabla^2\phi_B = 0$, meaning that the scalar potential obeys the Laplace equation.
The potential field solution is well understood and is typically given by a spherical harmonics expansion \citep{Altschuler1977}. 
Despite its simplicity, PFSS performs well in comparison to more elaborate MHD models \citep{Riley2006} and the low computational cost makes it one of the most widely used tools.

For our analysis we use the \emph{pfsspy} \citep{Stansby2020} implementation of the PFSS model. \emph{pfsspy} is an open source code which takes as input synoptic maps of the photospheric field, the value for the radial position of the source surface and the number of grid cells in the radial direction and then computes the full 3D magnetic field in the specified volume.  
Additionally, \emph{pfsspy} is fully integrated with \emph{astropy's} \citep{Astropy2018} coordinate and unit framework and has the capacity to trace magnetic field lines in the extrapolated volume. 
These features give us the ability to define the ballistically back-mapped point on the source surface in the Carrington coordinate frame (SkyCoord object) and then use it as a seed to trace its field line connectivity to the solar surface with \emph{pfsspy}. The resulting solar surface point is also expressed in the Carrington coordinate frame, which facilitates the plotting of these points on top of solar EUV observations. 
As input we used GONG synoptic magnetograms, these data are described in more detail in Sect. \ref{Noise_analysis}.

\begin{figure*}[h]
    \centering
    \includegraphics[scale=0.43]{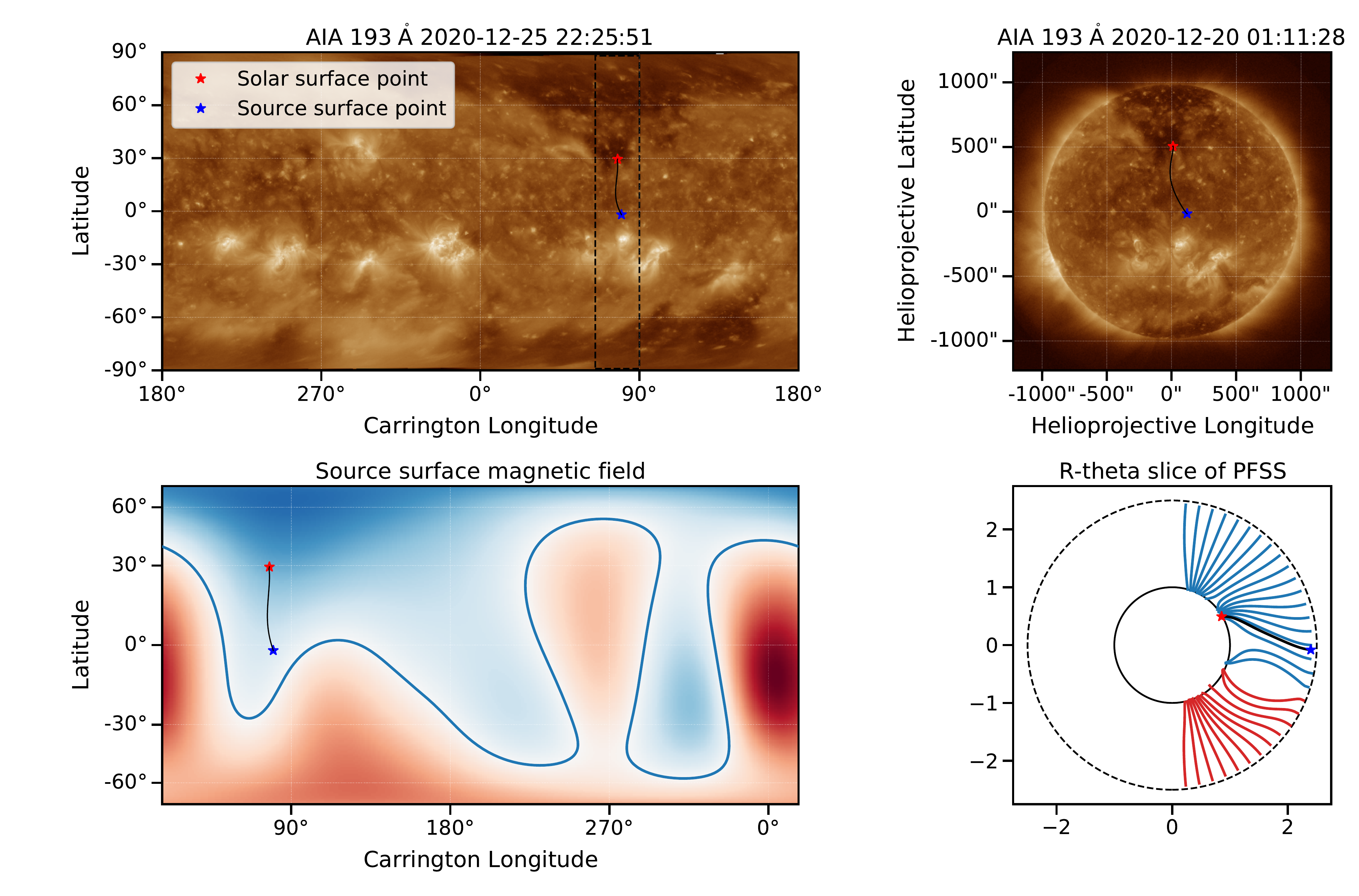}
    \caption{Overview plot of the back-mapping for a single in situ point (point number 10 of Event 1). The computed source surface point is represented with a blue star and the solar surface point with a red start for all the plots. The magnetic field line that connects the two points is displayed as a black line. (Top left) AIA 193 $\AA$ synoptic map. (Top right) AIA 193 $\AA$ full disk image at the time of the back-mapping. This image is the most accurate representation of the solar atmosphere at the time that is indicated with the dashed line box in the synoptic map. (Bottom left) The radial magnetic field at the source surface height. The blue line indicates the polarity inversion line and the colored regions the different polarities. (Bottom right) An R-theta slice of the 3D magnetic field that was computed from the PFSS extrapolation; the inner circle (solid black line) represents the solar surface and the outer circle (dashed line) the source surface, the blue and red line represent the magnetic field lines with different polarities.}
    \label{fig:overview_plot}
\end{figure*}

An overview plot of the back-mapping for a single in situ point (point 10 of Event 1) can be seen in Fig. \ref{fig:overview_plot}. The computed source surface point is represented with a blue star and the solar surface point with a red start for all the plots. For a visual representation of the magnetic mapping, the magnetic field line that connects the two points is displayed as a black line. The top left plot displays the AIA 193 $\AA$ synoptic map. The top right the AIA 193 $\AA$ full disk image at the time of the back-mapping. This image is the most accurate representation of the solar atmosphere for the time indicated with the dashed line box in the synoptic map. 
The radial magnetic field at the source surface height can be seen in the bottom left plot, in which the blue line indicates the polarity inversion line and the colored regions the different polarities. The bottom right plot is an R-theta slice of the 3D magnetic field that was computed from the PFSS extrapolation: the inner circle (solid black line) represents the solar surface and the outer circle (dashed line) the source surface, the blue and red lines represent the magnetic field lines with different polarities.

\subsection{Sources of uncertainty} \label{Uncertainty_sources}
The first step in an effort to define the uncertainty in the back-mapped position is to identify all the possible sources that can affect the final result. Sources of uncertainty can be different underlying assumptions of the framework, approximations, free parameters or sensitivity to certain conditions. 

Some of the input data of our framework are the in situ speed of the solar wind, the location of the spacecraft together with the clock time, and the synoptic magnetograms that are used to calculate the magnetic topology.
These input data have themselves intrinsic measurement errors, which can propagate in the final location of the source region.
The main free parameter comes from the magnetic mapping and it is the height of the source surface. 
This height is important because it affects strongly the configuration of the magnetic topology.
In almost all implementations of back-mapping with a PFSS model, a typical source surface height value of  2.5 \solrad{} is used but we will look at meaningful variations around this value.
Lastly, for the ballistic mapping the main assumption is the constant and radial propagation of the solar wind, with a typical error estimation of $10^{\circ}$, we explore if this can be improved by using more refined velocity profiles as discussed in Sect. \ref{Ballistic_Mapping}. 

\begin{figure*}[h]
    \centering
    \includegraphics[scale=0.35]{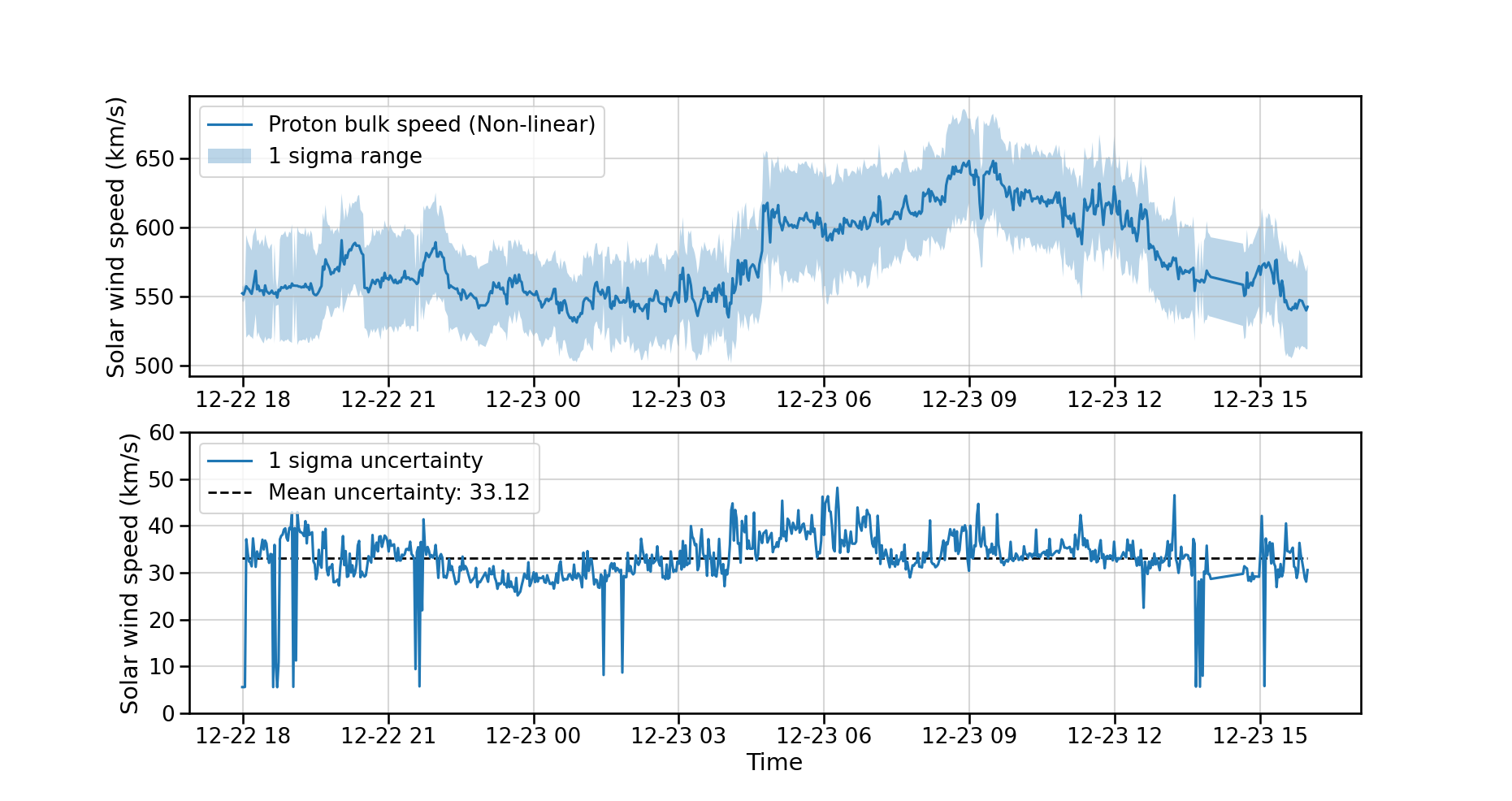}
    \caption{(Top) WIND measured solar wind speed (blue line) and its 1-sigma uncertainty as provided in the WIND data (shaded area). (Bottom) The 1-sigma uncertainty for the selected interval of our case study, together with its average value (horizontal dashed line).}
    \label{fig:sw_vp_1sigma}
\end{figure*}

The uncertainty in the position of the WIND spacecraft depends on the time at which the data are taken. During orbital maneuvers the uncertainty is tiny, on the order of a few tens of meters. The further from an orbital maneuver, the larger the uncertainty grows (e.g., due to accumulation of rounding errors and gravitational perturbations, etc).  At its largest, the uncertainty is on the order of 100 km.
Since the error in spacecraft position is very small, varying it yields negligible effect in the back-mapped location of the source region ($< 1^\circ$). 
Additionally, the error in solar wind speed as it is retrieved from the WIND data products and shown in Fig. \ref{fig:sw_vp_1sigma}, produces a very small uncertainty in the back-mapped position ($\leq 1^\circ$). This uncertainty is derived by applying the ballistic back-mapping procedure for the measured speed $\pm 1 \sigma$ and comparing the corresponding locations on the solar surface with that of the original measured speed.

Based on the above, in this study we will focus mainly on three sources of uncertainty: the choice of the velocity profile, the height of the source surface and the error in the input magnetograms. The effect of each uncertainty source on the final back-mapped location of the solar wind will be examined and a confidence interval will be computed. The analysis of each source is described in more detail in the following sections.


\subsubsection{Velocity profiles}\label{velocity_profiles}
In order to accurately sample the space of our `hybrid' velocity profiles (see Sect. \ref{Ballistic_Mapping}) we need a bound for the largest $r_s$ value, which comes from observational data.
As it is evident in Fig. \ref{fig:custom_profiles}, different studies of outflow speeds above coronal holes provide varying observational constraints. Not all observations are done in the same conditions, they include coronal holes at different latitudes, at different phases of the solar cycle, with varying morphology and different approximations in the Doppler dimming technique.
In this respect, the observational constraint we select depends on the properties of the solar wind parcel we want to back-map. 

In this study we are focusing on the back-mapping of fast solar wind flows measured at L1, which should have originated from a mid to low latitude coronal hole (Earth directed). With this in mind we can discard observations that are focused on polar coronal holes. 
UVCS data showed differences in the outflow velocity and temperature of the solar wind above equatorial and polar coronal holes, with the first presenting slower and cooler flows \citep{Miralles2004}, but in both cases similar in situ speeds. This can be seen in Fig. \ref{fig:custom_profiles} as all the lines that represent observational data from polar coronal holes use the star symbol. We note in passing that although we are focusing here on the fast solar wind, our methodology can easily be applied to slow solar wind regimes in a very similar fashion. 

The next step in narrowing down our selection of an observational constraint is the in situ measured solar wind speed. Due to the fact that UVCS can measure outflow speeds only off limb, there are very few cases where outflow speeds close to the Sun have been clearly associated with in situ speed measurements at larger distances. Most of these cases display an in situ solar wind of approximately 600 $\mathrm{km}\,\mathrm{s}^{-1}$ \citep{Miralles2004}, meaning that the validity of this observational constraint, for selecting the largest $r_s$ value, could potentially be challenged for much lower in situ solar wind speeds.
\citet{Miralles2004} provided a range of values for the outflow velocity above low latitude coronal holes, the lower bound of this range is the observational constraint which presents the biggest correlation with our data set (in situ speed) and is the one we use for our analysis (see the curve with circle markers in Fig. \ref{fig:custom_profiles}). Based on this we can selected the `hybrid' profile with the largest $r_s$.
This observational constraint, based on its very low speed values close to the Sun, will provide a first order bounding in the uncertainty of the source region location, which can be further be improved for specific cases of in situ solar wind. 

For our analysis, we sample uniformly the velocity profile space between the constant speed profile (dotted horizontal line in Fig. \ref{fig:custom_profiles}) and the largest $r_s$ profile, creating a sample of 20 profiles. The sampling in velocity profile space means sampling in the bmt space and by extension on the longitude that we arrive at the source surface. Next, we follow the magnetic topology to find the corresponding solar surface points. For the PFSS extrapolation a fixed source surface height of 2.5 \solrad{} was used.

\begin{figure}[h]
    \centering
    \includegraphics[]{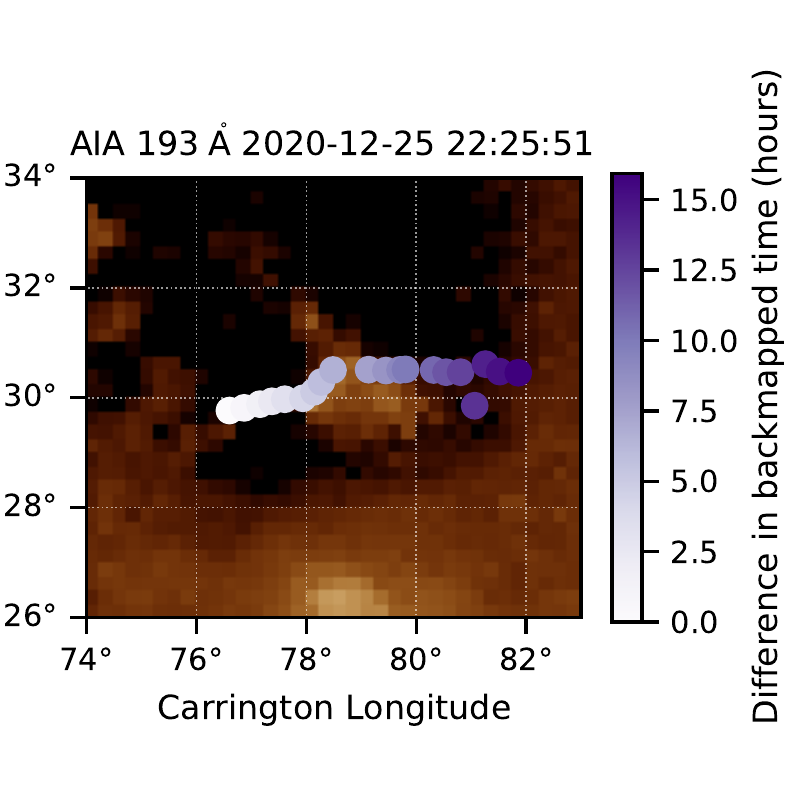}
    \caption{Back-mapped points on the solar surface derived from the velocity profiles analysis, for in situ point number 10 of Event 1. The background is a zoom-in of a synoptic AIA 193 \AA \, map and their color represents the difference in their back-mapped time compared to the constant speed point (smallest travel time). 
    }
    \label{fig:v_profiles}
\end{figure}

Results for varying across the 20 velocity profiles for the in situ point number 10 (indicated in Fig. \ref{fig:sw_v-d_ev4}) are shown in Fig. \ref{fig:v_profiles}.
The derived solar surface points are plotted on a synoptic EUV map, that was created with full disk 193 \AA \, images from the Atmospheric Imager Assembly (AIA; \citealt{Lemen2012}). The displayed region is a zoom-in of the full synoptic map at the coronal hole that generated the observed wind. The low intensity areas of the map (seen as black) represent the coronal hole. 
In this figure, the solar surface points for the 20 velocity profiles considered are displayed as colored circles (the 20 circles actually overlap and stay rather close to the constant speed estimate). Their color represents the difference in back-mapping time compared to that of the constant speed profile point, as shown in the adjacent colorbar. Additionally, the solar surface point of the constant speed profile (shortest travel time) is displayed with a red star and the solar surface point of the largest-$r_s$ profile (largest travel time) with a cross.



\subsubsection{Source surface height} \label{SS_height}

The height of the source surface in the PFSS models can have significant effect on the magnitude of the magnetic field and the amount of open flux, which in turn will affect the shape of the Heliospheric Current Sheet (HCS). 
The canonical value of the source surface height that is widely used during PFSS extrapolations is 2.5 \solrad{} and it is based on earlier works by Hoeksema, Wilcox, and Scherrer (1982, 1983) and Hoeksema and Scherrer (1986). 

But as discussed previously this is a free parameter and it is reasonable to assume that the height of the SS varies due to different factors, such as with the solar cycle. 
\citet{Lee2011} showed that during minimum solar activity lower SS heights generate results that better match the observations, with values between 1.5-1.9 \solrad{} being the optimal fit. 
There are also strong indications that the SS deviates from the purely spherical shape, with its height depending on longitude and latitude  \citep{Schulz1978,Levine1982,Panasenco2020,Kruse2020}. 
\citet{Schulz1978} introduced first a non-spherical source surface description, which was tailored initially for bipolar and later on for quadrupolar magnetic fields, but under more realistic boundary conditions with multipolar magnetic fields this description presents computational challenges \citep{Levine1982,Schulz1997,Lee2011}.
More recently, \citet{Panasenco2020} reconstructed a  non-spherical source surface over a period of 6 months, from a time series of spherical SS heights that, each, best matched the polarity inversions observed by Parker Solar Probe. They showed that the source surface height tends to be lower above polarity inversion lines, and that it typically lays between 1.8-2.5 \solrad{}, although it can reach values as low as 1.2 \solrad{} in certain locations.
Whereas \citet{Kruse2021}, with the investigation of elliptical source surfaces for the PFSS extrapolation, found indications that during solar minimum an oblate elliptical SS performs better than the spherical SS, resulting to equatorial heights of around 3 \solrad{}.


Due to the variation of the SS height (either along the solar cycle or with magnetic features present on the solar surface), it is important to quantify its effect on the back-mapped position of the solar wind and identify a confidence interval. 
For this reason, we work in a similar fashion as before (Sect.\ref{velocity_profiles}) and sample the SS height space, deriving a number of solar surface locations for a single in situ point. For the ballistic portion of the back-mapping the norm of a constant speed profile is used to arrive at the source surface.
For our study we have selected a range of source surface heights between 1.5-3.5 \solrad{}, as it encompasses all the acceptable values that have been reported.

\begin{figure}[h]
    \centering
    \includegraphics[]{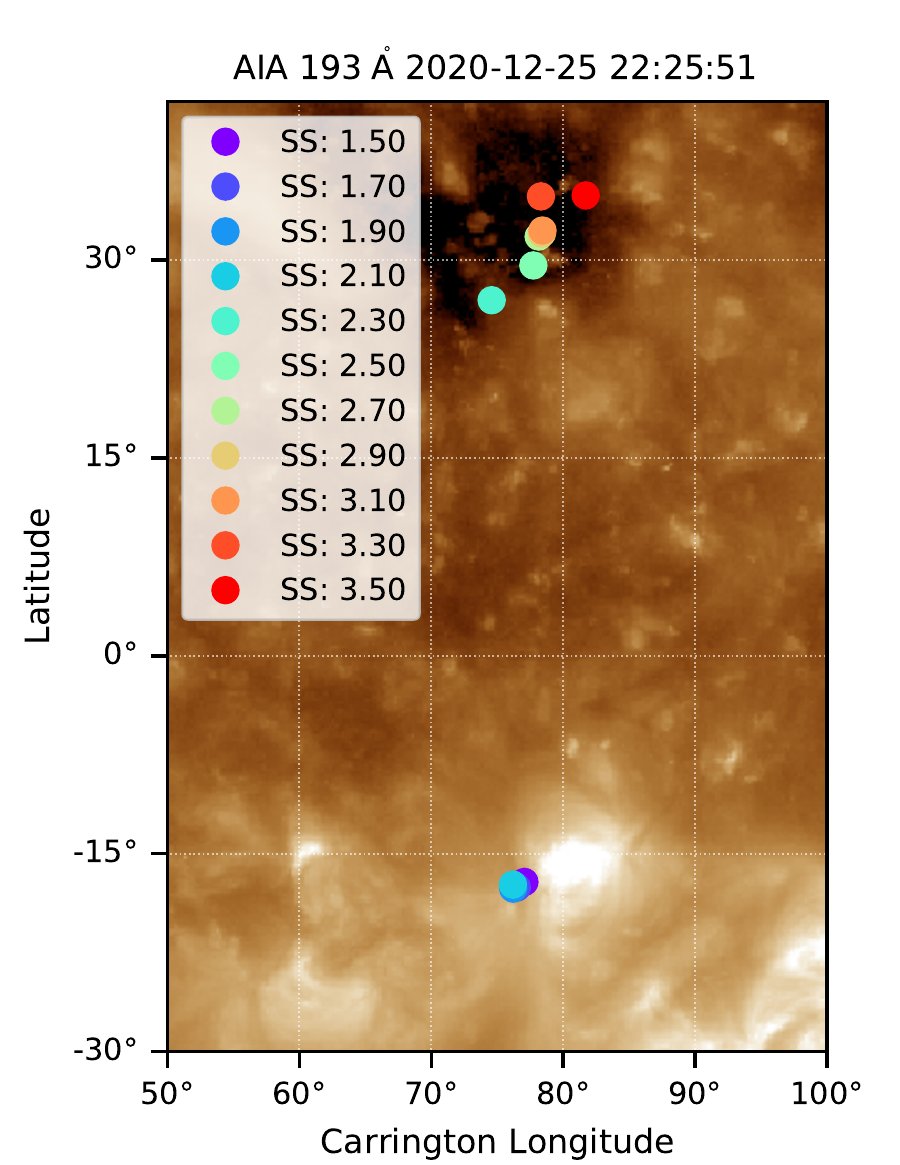}
    \caption{Back-mapped points on the solar surface derived from varying the source surface height for in situ point number 10 of Event 1. The points are plotted on a zoom-in of the synoptic AIA 193~$\AA$ map as background. The color of the solar surface points represents the source surface height that was used during their magnetic mapping.}
    \label{fig:ss_scatterplot_euvmap}
\end{figure}

Results of this analysis for in situ point number 10 of Event 1 and with a sample size of 11 SS heights, can be seen in Fig. \ref{fig:ss_scatterplot_euvmap}. The solar surface points are presented with circular markers on top of a zoom-in region of a synoptic AIA 193 \AA \, map (similar to Fig. \ref{fig:v_profiles}). The color of the marker represents the SS height that was used to derive this point. 
As we transition from lower to higher SS heights a clear shift in the connectivity is observed. For lower heights ($\leq 2.3~$\solrad{}) the back-mapped points are traced to a location close to the equatorial active region, but for higher heights ($>2.3~$\solrad{}) we get a clustering to the low latitude coronal hole.   

This can be interpreted as the following: because the wind is first back-mapped to the source surface at the latitude of the equator, it is more likely that field lines reaching the low SS heights correspond to those that come from the edges of the AR, which is  closer to this same back-mapped point. In other words, as the SS height goes lower, higher order multipoles present in the AR fields are able to intersect it.

\subsubsection{Magnetogram noise} \label{Noise_analysis}
The synoptic photospheric magnetograms that are provided as input in the PFSS model are a crucial component of the back-mapping framework as they have a direct impact in the calculated magnetic topology of the low corona. 
Therefore the sensitivity of the magnetic topology on the boundary conditions (synoptic magnetograms) will provide an uncertainty in the back-mapped position on the solar surface. In order to quantify this sensitivity we add noise to the input magnetograms and calculate the effect on the source region location. As mentioned before, the photospheric synoptic magnetograms that we use for our analysis are taken from GONG.

The Global Oscillation Network Group \citep{Harvey1988,Harvey1996,Leibacher1999}, is a community-based program that was designed to study the conditions in the solar interior using the information of acoustic waves that propagate through the Sun (helioseismology).
To accomplish that, GONG developed a network of six identical instruments around the world, to record the Doppler velocity of the solar surface and thus obtain nearly continuous observations of the Sun's "five-minute" oscillations, or pulsations. Despite the fact that magnetic field measurements were not the primary objective in the design of GONG they have become one of the most important and widely used products, with 75\% of all the publications that cite the GONG instrument paper \citep{Harvey1996} using the GONG data for space weather related global field extrapolations \citep{Plowman2020a}. After the latest updates in the GONG network (GONG+ and GONG++) high quality magnetograms are obtained every minute at each site of the network \citep{Hill2008}.

For our analysis we need to determine the noise level in the GONG synoptic magnetograms. This noise level is found to be 0.5 G with three independent methods. The first, and the most important, comes from the underlying processes that produce the synoptic map and the GONG characteristics. The other two are based on noise detection techniques.
The GONG network is in operation since 1995 and had many upgrades through the years, making the identification of the appropriate noise level for synoptic maps not straightforward. 
\citet{Hill2008} and \citet{Harvey2009} quote a noise level of 3 G per pixel, but this noise refers to a single full disk magnetogram (1-minute cadence) and is an average value per pixel near the disk center. The noise would be higher for pixels closer to the solar limbs. 
Normally, 1-minute magnetograms are averaged to produce 10-minute full disk magnetograms. This is the main data product for GONG magnetic field measurements. 
Files for 10-minute full disk magnetograms include standard deviations for each pixel. These can be used to estimate the uncertainties for each synoptic map pixel, which represent both errors in measurements, and real variations, e.g. evolution of magnetic flux in each pixel. 
These uncertainties result in an approximately 0.5 G (0.42-0.5) noise level for a synoptic map pixel near the disk center, and this value is expected to increase when going towards the solar limbs (Alexei Pevtsov; private communication). 
We additionally confirm this 0.5 G noise level of the synoptic magnetogram with two noise detection methods. The first is based on wavelet analysis \citep{Donoho1994} and is implemented with the Python package \emph{scikit-image} \citep[Module:restoration, function:estimate\_sigma]{VanDerWalt2014}. 
The second is based on the work of \citet{Immerkaer1996}, which uses the Laplacian of the image to derive an estimate for the noise level. 
The calculated noise estimate from these functions, for synoptic GONG magnetograms, is $\approx$ 0.4-0.6 G.



\begin{figure*}[h]
    \centering
    \includegraphics[width=\textwidth,scale=0.64]{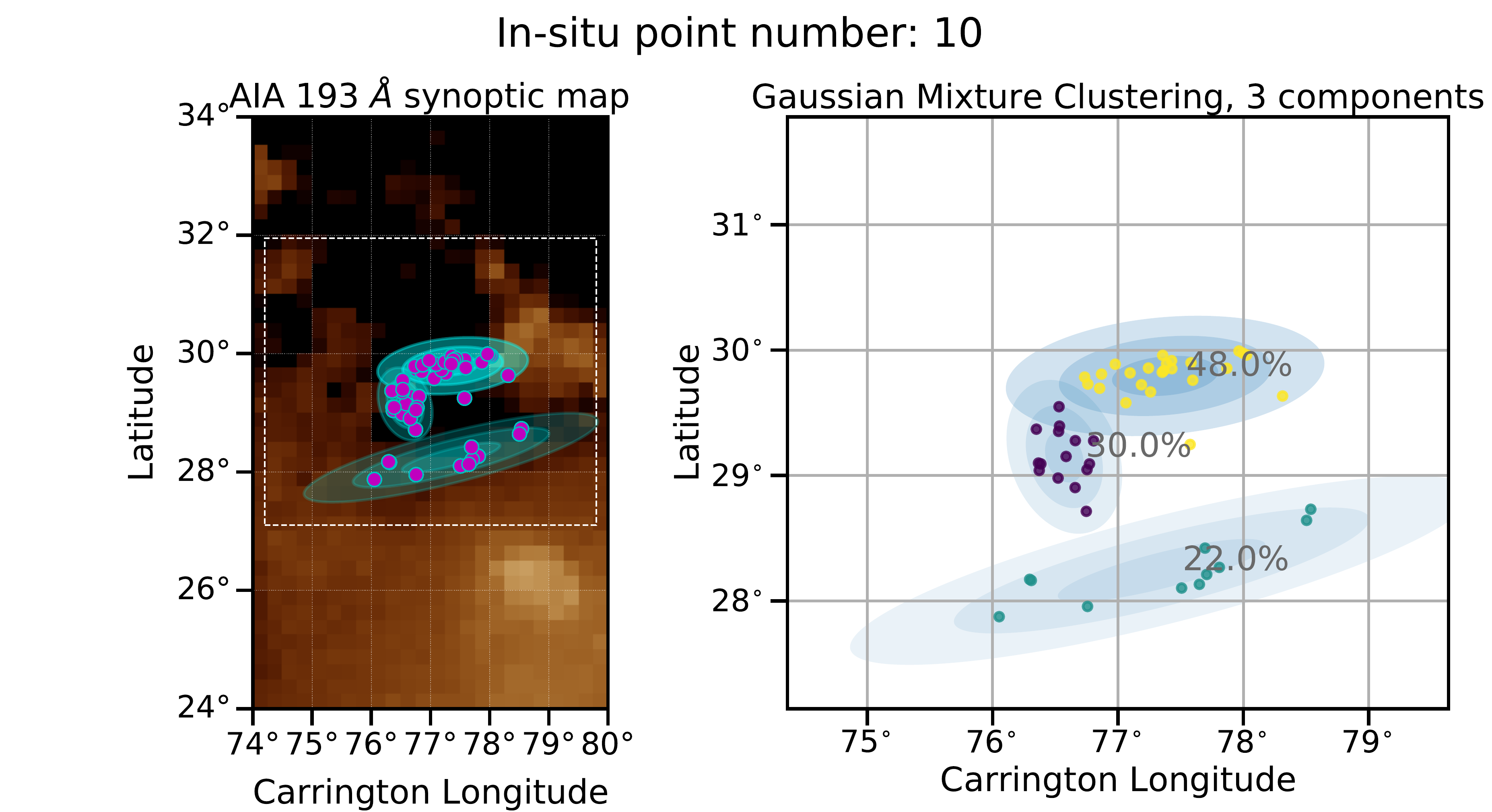}
    \caption{(Left) Back-mapped points (purple circles) on the solar surface, as derived from the Monte Carlo simulation of input magnetogram noise, for in situ point 10 of Event 1. 
    The back-mapped points are plotted on a zoom-in of the synoptic AIA 193~$\AA$ map. The white dashed box delimits the area that is enlarged in the right panel (without the EUV background image). (Right) Gaussian Mixture clustering of the back-mapped solar surface points from the Monte Carlo simulation. Here the optimal number of cluster was found to be 3. The light blue ellipses represent the confidence intervals of each cluster and next to them the probability of connection to each cluster is displayed. The level of transparency of each ellipse displays the variance interval they portrait, meaning the least transparent represents the 1$\sigma$ confidence interval and the more transparent the 3$\sigma$ confidence interval. These are also overplotted in the left panel.}
    \label{fig:noise_euv_plot}
\end{figure*}

Based on the above, 0.5 G is selected as an appropriate noise estimation in the input magnetograms, and in order to examine its effect on the back-mapped position at the solar surface we perform a Monte Carlo simulation. 
The steps of this simulation are the following.
First, we select a set of random noise values from the normal distribution (0,0.5), with mean 0 and standard deviation 0.5, and add it to the pixels of the input magnetogram, meaning that each pixel has a different (random) noise value.
Next, we perform the magnetic extrapolation using \emph{pfsspy} and having the source surface height set at the canonical value of 2.5 \solrad{}. 
Lastly, a specific point at the source surface is traced through the newly computed magnetic topology to the solar surface. This source surface point is derived from the ballistic mapping (constant speed) of a single in situ point and remains the same for the whole simulation.
This way, a single in situ measurement of the solar wind speed is correlated to a number of solar surface points, the spread of which represents the uncertainty in the source location originating from the magnetogram noise. 

We examined the effect that a different number of Monte Carlo runs ($n_{MC}$) have on the derived solar surface points. We found that, in general, above $n_{MC} = 80$ we do not have any more significant changes in the uncertainty area. Furthermore, our analysis showed that in some cases even from $n_{MC} = 50$ we have sufficient convergence.


Results of the magnetogram noise analysis for the example of in situ point number 10 of Event 1 are seen in the left panel of Fig.~\ref{fig:noise_euv_plot}. 
Each circular marker (purple circle) represents the solar surface point that was derived for one realization of noise for the input magnetogram in the Monte Carlo simulation. 
The solar surface point that was computed without added noise in the input magnetogram is displayed with a red star marker. A sample of 50 noise realizations was used in this example. 
The background image, similar to Fig. \ref{fig:v_profiles} and \ref{fig:ss_scatterplot_euvmap}, is a zoomed-in region of the corresponding AIA 193 $\AA$ synoptic map.
Although the synoptic EUV map does not represent the morphology of the coronal hole at the time correlated to the back-mapping time, it provides a good approximation to inspect the association of the back-mapped points with the coronal hole.

In order to investigate in depth the grouping of the back-mapped points we use the Gaussian Mixture Model (from here on as GMM).
A Gaussian mixture model is a probabilistic model in which all data points are assumed to be generated by a mixture of a finite number of Gaussian distributions with unknown parameters. 
Mixture models can be thought of as a generalization of simpler clustering techniques that provide a fixed convex shape for the clusters, by including information about the covariance structure of the data as well as the centers of the latent Gaussian distributions.
The Python package \emph{scikit-learn} \citep{Pedregrosa2011} implements various classes for estimating Gaussian mixture models, each of which corresponds to a different estimation strategy. 

For this study we focus on the \emph{GaussianMixture} module\footnote{https://scikit-learn.org/stable/modules/mixture.html}, which implements the expectation-maximization algorithm for fitting mixture-of-Gaussian models. 
The main input parameters of this model is the number of clusters and the type of covariance matrix.
The optimal number of clusters, typically referred to as number of components, is a crucial parameter in all clustering algorithms and cannot be derived in advance with absolute certainty. One way to estimate the optimal number of components is based on the inspection of information criteria such as the Bayesian Information Criterion (BIC) and the Akaike Information Criterion (AIC). In \emph{scikit-learn} these are expressed as
\begin{eqnarray}
    BIC & = & -2 \log(\hat{L}) + \log(N)d \,,\\
    AIC & = & -2 \log(\hat{L}) + d \,,
\end{eqnarray}
where $\hat{L}$ is the maximum likelihood of the model, $d$ is the number of parameters and $N$ the number of samples.
We investigated these criteria together with another model, the Bayesian Gaussian Mixture, but a definitive and automated way to derive the optimal number of components could not be achieved. For this reason we created a custom function that identifies the optimal number of clusters based on two parameters, the overlapping of neighboring clusters and the minimum number of elements allowed in each cluster. 
For the other input parameter, we have selected the `full' covariance type. This allows each component to have its own covariance matrix. Essentially meaning that for every component, a confidence ellipsoid can be drawn with a shape that is independent of the other components.


The results of the GMM clustering are presented in the right panel of Fig.~\ref{fig:noise_euv_plot}. 
Similar to the left panel, each circular marker represents a back-mapped position on the solar surface for one noise realization (in the input magnetogram).
The different colors of the markers, indicate the cluster that they belong.
Additionally, we can compute an uncertainty area for every cluster, as shown by the ellipses in Fig.~\ref{fig:noise_euv_plot}. The level of transparency in each ellipse represents the confidence interval that it displays. Here the 1-3 $\sigma$ intervals are shown, with the least transparent ellipse representing the 1 $\sigma$ confidence interval.
These ellipses can be displayed on the synoptic EUV map (left panel of Fig. \ref{fig:noise_euv_plot}) or on a full disk EUV image providing us with a confidence area for the source region of the solar wind parcel that we measure in situ. This confidence area represents the uncertainty coming from the sensitivity of the extrapolation to the boundary conditions and can be computed for other sources of uncertainty as well.

\subsubsection{Uncertainties aggregation}
\begin{figure*}[h]
    \centering
    \includegraphics[]{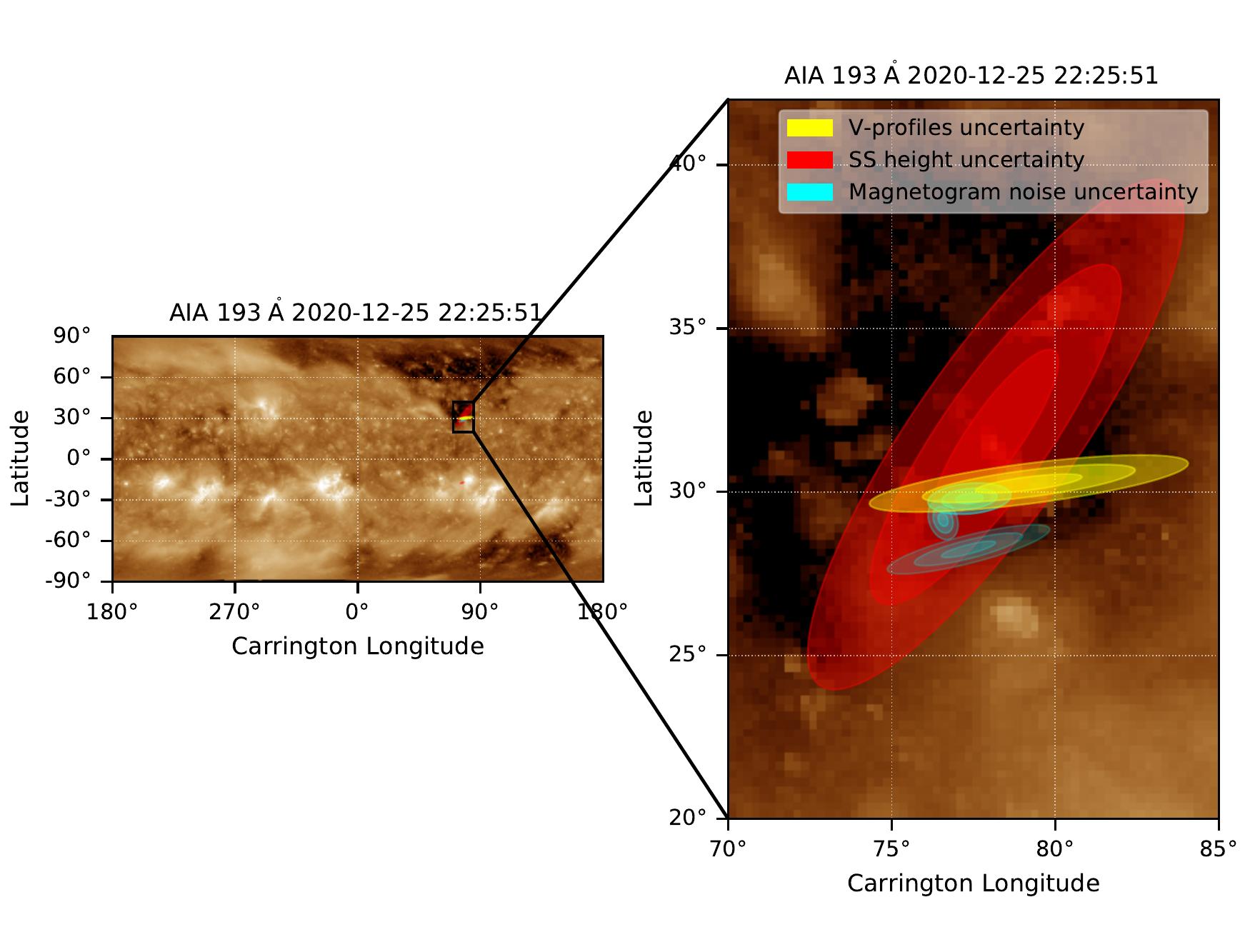}
    \caption{Confidence areas for the source region of the solar wind derived for different sources of uncertainty. (Left) The synoptic AIA 193~\AA{} map corresponding to Event 1. The black box indicates the area of interest. (Right) A zoom-in of the area of interest. 
    The colored ellipses display the confidence area in the back-mapped position (on the solar surface) of the in situ solar wind for different sources of uncertainty. 
    The color of each ellipse denotes the uncertainty source responsible (as indicated in the legend) and the transparency the 1-3 sigma area, with the 1 sigma being less transparent. These confidence area results correspond to in situ point number 10 of Event 1.}
    \label{fig:all_sources_euv_plots}
\end{figure*}

The process that was used to derive the uncertainty area in the case of the magnetogram noise can be applied to the other sources of uncertainty as well. The results for the three uncertainty sources that we examined can be seen in Fig. \ref{fig:all_sources_euv_plots}.
The 3 sigma confidence ellipses for the three sources of uncertainty are presented in a single figure. The left plot displays the AIA 193 $\AA$  synoptic map together with a dashed box for the region of interest. The right plot is an enlarged view of the region of interest. The uncertainty in the back-mapped location is displayed with the colored ellipses. The color of the ellipses indicates the source responsible for this uncertainty and their transparency the 1-3 sigma area, with the 1 sigma being less transparent. 
This view clearly shows the impact of each source of uncertainty, with the height of the source surface having the most significant effect, followed by the uncertainty in the velocity profiles and the magnetogram noise. 

We assumed here that the three uncertainties behave independently one from another. 
As this assumption is only a first order approximation, a proper aggregation of the uncertainties would require to cluster a number of Monte Carlo realizations produced by randomly selecting, for each realization, a velocity profile ($r_s$), a source surface height and a magnetogram with random noise. But this is out of the scope of the current study, as our aim is to illustrate the different sources of uncertainty and how they compare one to another.

\subsection{Statistical analysis on multiple events} 
From the analysis of a single in situ point we can derive an estimation about the effect of each source of uncertainty in the source region of the solar wind and compute a confidence area. 
To generalize these local, single-point estimations we need to inspect the connectivity of multiple in situ points and also multiple high speed streams. 

\begin{figure*}[h]
    \centering
    \includegraphics[scale=0.34]{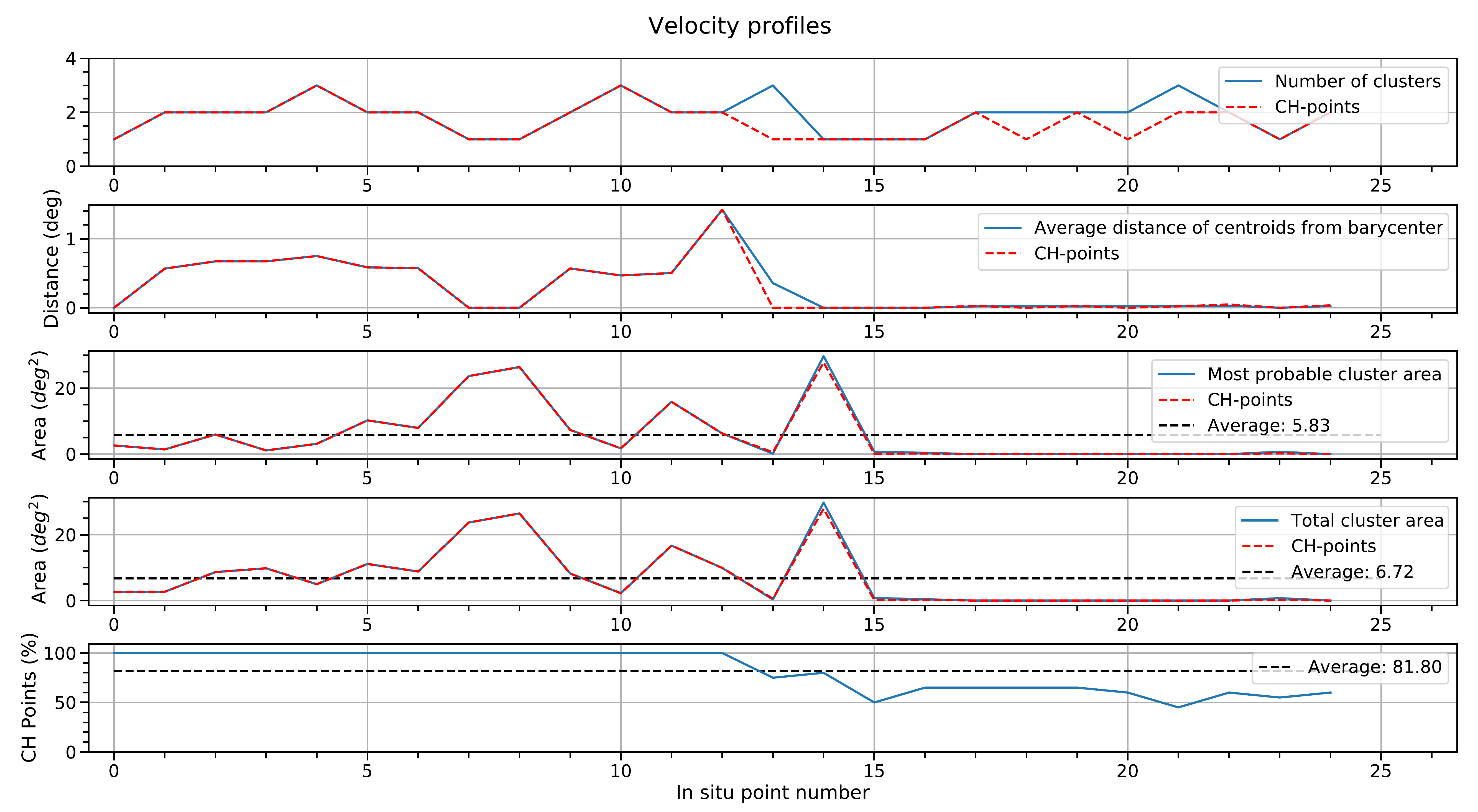}
    \caption{Evolution of 4 metrics that we use to assess the quality of our back-mapping, for the velocity profiles source of uncertainty, of all the in situ points of Event 1. The horizontal axis indicates the in situ points number (selected solar wind measurements) of the Event 1 high speed stream, as indicated previously in the inset plot of Fig.~\ref{fig:sw_v-d_ev4} (red stars markers). From top to bottom: the optimal number of clusters that was found, the average distance of each cluster centroid to the barycenter of the clusters, the uncertainty area of the most probable cluster, the total uncertainty area, and the percentage of the back-mapped points that ended up inside the coronal hole (as identified from the AIA 193 $\AA$ passband). 
    For the first 3 metrics we present the values calculated for all the back-mapped points, as indicated with the blue line, and the values for only the back-mapped points that are located inside the coronal hole, as the dashed red line.}
    
    \label{fig:ev4_stats_vprof}
\end{figure*}

First, we focus on the quantities that can be computed for all the in situ points of a specific high speed stream, in this case the fast solar wind interval that is shown in Fig. \ref{fig:sw_v-d_ev4} (Event 1).
These quantities are computed for each source of uncertainty separately. The first one is the optimal number of clusters that was found with the Gaussian Mixture Model, which provides an indication of how closely correlated are the back-mapped points and in particular if, when perturbing the initial value of our free parameters, we fall on different sides of a separatrix in the magnetic topology. Since the number of clusters can display only one aspect of the correlation we have also computed the barycenter of all the clusters and calculated the average distance from each cluster centroid to the barycenter. This metric works as an indicator for the spatial proximity of the clusters. Next, we compute the 3 sigma uncertainty area from the confidence ellipses of each cluster and examine the area of the most probable cluster and the total area from all the clusters. Lastly, the percentage of back-mapped points that ended up inside the coronal hole, as this was identified from the AIA 193 $\AA$ images by using a thresholding technique, is calculated.
An example of this analysis, for the uncertainty in the velocity profiles, can be seen in Fig. \ref{fig:ev4_stats_vprof}. 

\begin{figure*}[h]
    \centering
    \includegraphics[scale=0.59]{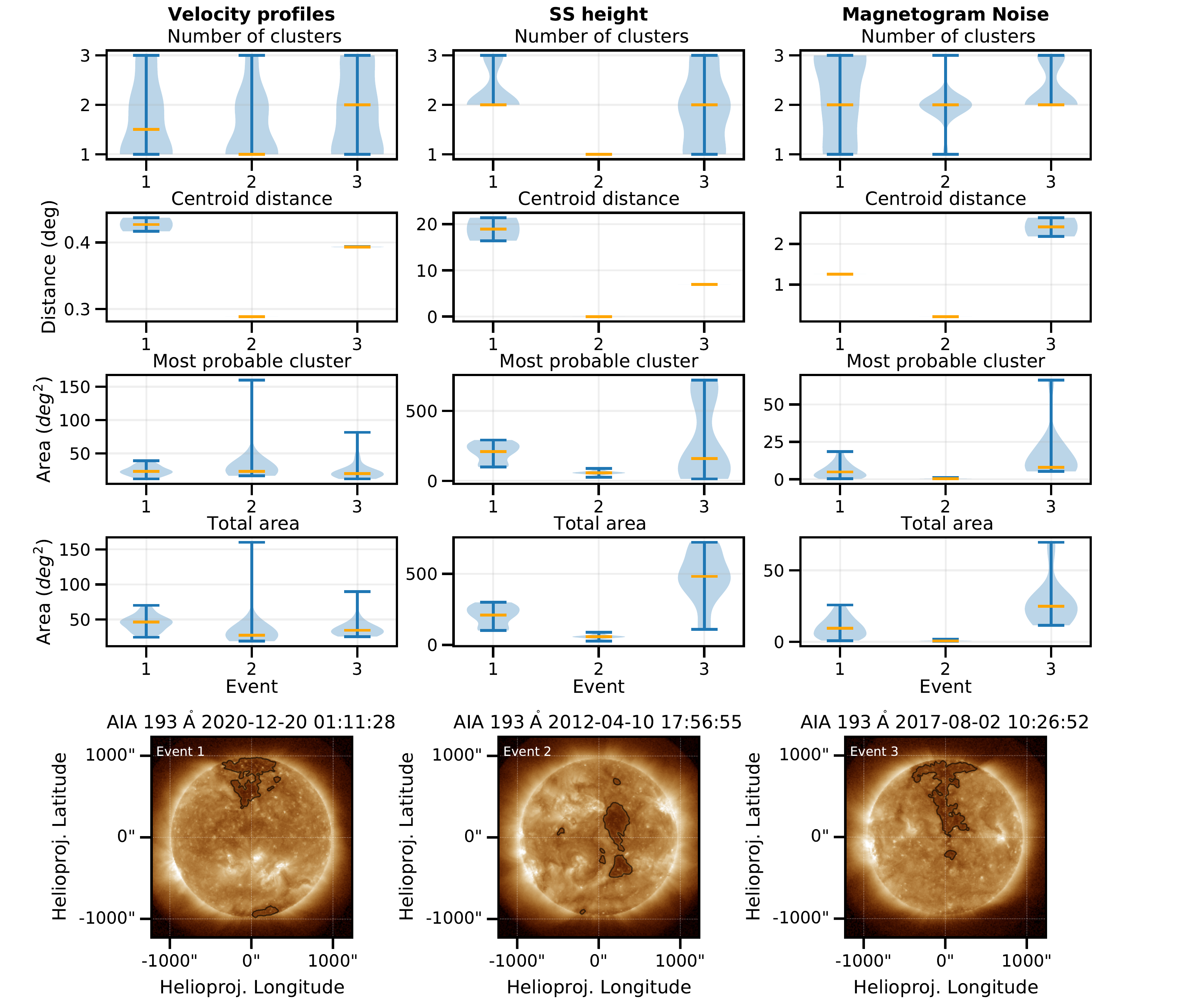}
    \caption{Violin plots (first four rows) of the back-mapping assessment metrics, for every source of uncertainty in Events 1, 2 and 3. The horizontal axis represents the event (high speed stream) that was examined. Each column corresponds to a source of uncertainty and each row to a statistical quantity: optimal number of clusters, average distance of cluster centroids from the barycenter, the 3-sigma uncertainty area of the most probable cluster and the total 3-sigma uncertainty area. In each violin plot, the median values are indicated with an orange line and the extreme values with light blue lines. Where the violin plot is not visible, the distribution is very concentrated around a single value. The last row displays AIA 193 $\AA$ images, taken approximately at the midtime of each event, including the contours of the coronal holes present at the time. Event 1 and 3 are considered as polar coronal hole extensions and Event 2 as a low latitude coronal hole.}
    \label{fig:ev4_6_7_all_stats}
\end{figure*}

Naturally, our framework and analysis can now routinely be done for many more events of interest. To prove this point, we analyzed two more high speed streams on 13-14/04/2012 and on 4-5/08/2017, associated with a low latitude coronal hole and a polar coronal hole extension, respectively. We will refer to these as Event 2 and 3 from now on, considering Event 1 the case study high speed stream from Fig.~\ref{fig:sw_v-d_ev4}, more details for these events can been seen in Table \ref{table:1} and a context image of their morphology in the bottom row of Fig.~\ref{fig:ev4_6_7_all_stats}.
The methodology for these events follows exactly what we displayed so far.
The results collected for all three events are shown in the form of violin plots in Fig. \ref{fig:ev4_6_7_all_stats}. A violin plot depicts distributions of numeric data for one or more groups using density curves (or kernel density estimate (KDE)). The width of each curve corresponds with the approximate frequency of data points in each region, this way an overview of the distribution for multiple groups can be presented in a single figure. 
In Fig. \ref{fig:ev4_6_7_all_stats} each column of the violin plots represents the source of uncertainty that was examined and each row one of the first four quantities from Fig. \ref{fig:ev4_stats_vprof}. The horizontal orange line in the violin plots indicates the location of the median value and the two light blue horizontal lines the location of the max and min values. Where the violin plot is not visible, the distribution is very concentrated around a single value. 
In the last row of Fig.~\ref{fig:ev4_6_7_all_stats} three AIA 193 $\AA$ full disk images are displayed, taken approximately at the midtime of each event. The corresponding event number for each image is indicated in the upper left corner of the images. Additionally, we have overplotted the contours of the coronal holes in every AIA image, as computed from the method described below in Sect.~\ref{framework_evaluation}.

The back-mapping uncertainties, as shown in the violin plots of Fig.~\ref{fig:ev4_6_7_all_stats}, are strongly influenced by the coronal hole morphology of every event. Noticeably, for all the back-mapped points of Event 2, the back-mapping process yields a compact source region of the solar wind that remains relatively unaffected by the different perturbations: all perturbed solutions are grouped within a very small area (as seen in Fig.~\ref{fig:ev4_6_7_all_stats} from the small centroid distance between the clusters and the total area). This is clearly associated with the size, location and shape of the coronal hole, which presents a significant extent both in longitude and latitude, is located near the equator (where the spacecraft also lies) and is at the same time more compact and less fragmented than the other coronal holes of our study. As a result, the magnetic topology is very simple and consists in field lines that are mostly radial, with no major topological discontinuity that could be crossed when the initial conditions are perturbed. 
The opposite effect can be seen in event 3, where a polar coronal hole extends all the way to the equator, creating a very big area of possible back-mapped locations. This is especially true in the case of the SS height uncertainty: as the SS height increases, connections to ever larger latitudes occur, resulting in a very big confidence area for the source region of the solar wind.

After inspection of Fig.~\ref{fig:ev4_6_7_all_stats}, it is again evident that the height of the source surface produces the biggest uncertainty in the back-mapped location, followed by the uncertainty in the velocity profiles and the noise in the input magnetogram. This is a strong indication for the hierarchical significance of each source of uncertainty.

\subsection{Framework evaluation}\label{framework_evaluation} 
Unfortunately, we do not know with absolute precision the area on the Sun from where the in-situ measured solar wind originates. This lack of ground truth makes it difficult to asses which back-mapped locations, derived after perturbing different components of the back-mapping framework, can be considered as reasonable indications of the solar wind source region. Consequently, it is important to try to evaluate our back-mapping results with other methods. One of these methods is the comparison with the results of another, well established, back-mapping framework. This is the Magnetic Connectivity Tool \citep{Rouillard2017,Rouillard2020}. 
 
\begin{figure*}[h]
    \centering
    \includegraphics[]{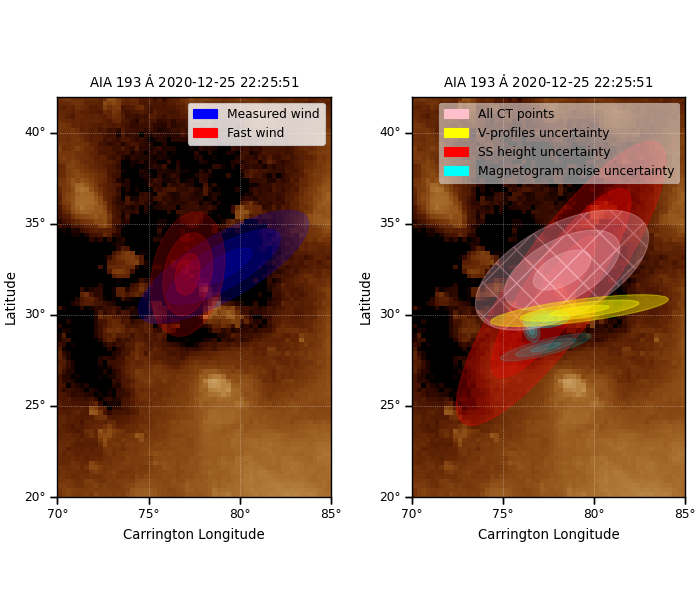}
    \caption{Uncertainty area of the solar wind source region after clustering the Magnetic Connectivity Tool results for in situ point 10 of Event 1. The background is a zoom-in of the synoptic AIA 193~\AA\, map that corresponds to this observation. (Left) The uncertainty area for the measured solar wind data is indicated with blue and the uncertainty area for the fixed solar wind values is indicated with red. (Right) The uncertainty area of both fast and measured solar wind combined is indicated with the hatched pattern ellipse, superimposed on our own uncertainty areas for each source, as shown in Fig. \ref{fig:all_sources_euv_plots}.}
    \label{fig:ct_clusters_euv_plot}
\end{figure*}

The Gaussian mixture clustering can also be applied to the existing Magnetic Connectivity Tool results.
The Magnetic Connectivity Tool uses in situ measurements of the solar wind but when these are not available to the framework it returns 2 groups of results, one with a fixed value of 800~\kms (assuming fast wind) and one for 300~\kms (slow wind).
For the purpose of comparing with our own framework, we can cluster independently the "measured solar wind velocity" results, blue contours in the left panel of Fig.~\ref{fig:ct_clusters_euv_plot}, and the results assuming the default value for the fast solar wind, red contours. On the right panel of Fig.~\ref{fig:ct_clusters_euv_plot}, the clustering of the combined wind measurements is displayed on top of the confidence areas for the different uncertainty sources that we computed. The background is the zoomed-in region of the AIA 193 $\AA$ synoptic map, similar to the right plot of Fig. \ref{fig:all_sources_euv_plots}. 
By comparing the connectivity of the Magnetic Connectivity Tool with the results of our framework we can see that they are almost co-spatial. 
The differences lie mostly in the shape and orientation of the derived uncertainty area, with the ellipses of the source surface height from our framework being more extended in the north-south direction due to the wide range of SS heights that was taken into consideration.

An additional metric that can be used to evaluate our framework is the correlation of the back-mapped points with the coronal hole, as it is seen in the EUV images and specifically in the AIA 193 \AA \ passband. This is applicable because we examine fast solar wind streams which should originate from coronal holes. The coronal hole boundaries are extracted after masking the original image with a certain threshold value (25$\%$ of the mean pixel value in the image) and applying a 2D Gaussian smoothing function to the data. This function is taken from the Python \emph{scipy} package \citep{Virtanen2020} and particular the \emph{ndimage} library. An example of this metric for the velocity profiles uncertainty can be seen at the bottom row of Fig. \ref{fig:ev4_stats_vprof}. 
On average we observe that the percentage of the back-mapped points that are located inside the coronal hole is around 85\%. Additionally, when we inspect the points that are outside the coronal hole boundaries, we see that they are located, most of the time, at the edge of the coronal hole or very close to it. To study this connection further we examine the connectivity of a few additional high speed streams where the coronal hole from which the solar wind originated was very small or patchy. The outcome of this analysis was that the coronal hole percentage of the back-mapped points was decreasing with patchier coronal holes. This outcome is expected and associated with the underlying assumptions of the PFSS model, which are discussed in more detail in Sect. \ref{Discussion}. 

\section{Discussion}\label{Discussion}

This work presents a framework for the back-mapping of the fast solar wind and the estimation of the uncertainty in the derived location on the solar surface. The performance of this framework is in agreement with a similar back-mapping tool (Magnetic Connectivity Tool), the results follow what is expected from the underlying physical processes, and confirm that the height of the source surface produces the biggest uncertainty.

The main limiting factor of this framework and other similar back-mapping processes is the extrapolation method. PFSS is known to be a less reliable approximation in areas of complex magnetic topology such as active regions and in general during the maximum of the solar cycle, where the coronal magnetic field is further away from a potential state. On the other hand, low latitude coronal holes that can produce high speed streams observable from Earth appear mostly in periods of increased solar activity. 
This creates some inaccuracy in the magnetic field topology. The alternative would be the use of MHD models, but such models are quite computationally expensive and can not be run routinely. 
As PFSS is the most widely used method to derive the magnetic field topology for solar wind back-mapping, a precise estimation of all the uncertainties associated with it and their impact is very important for the study of the solar wind.

As mentioned in Sec. \ref{velocity_profiles} this back-mapping framework can be extended to slower solar wind speeds, by adapting our selection of observational constraints in the velocity profile space. Additionally, it can be extended to use data from many other spacecrafts, not only WIND. This is possible by retrieving the spacecraft’s instant location using the \emph{sunpy} library \citep{Mumford2015} (in any coordinate system) and converting it to Carrington coordinates, which are then used for the back-mapping. That is particularly important in the current era of multiple missions away from Earth, like Parker Solar Probe and Solar Orbiter. 

\begin{figure}[h]
    \centering
    \includegraphics[scale=0.85]{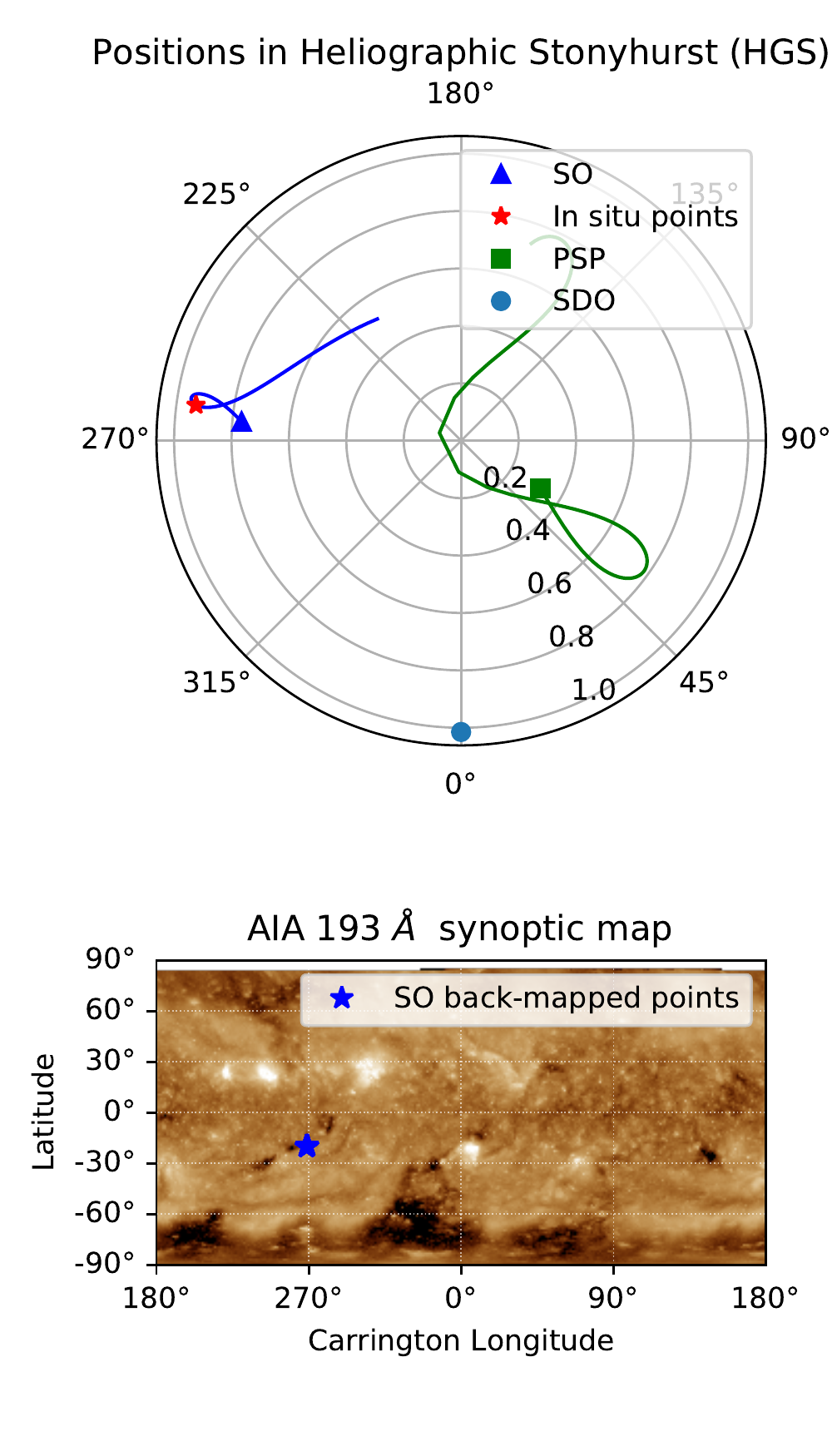}
    \caption{(Top) Solar Orbiter orbit (blue line) in the Heliographic Stonyhurst frame together with the orbit of Parker Solar Probe and the location of the Solar Dynamic Observatory. The orbits have been plotted within $\pm 80$ days around the time of the high speed stream observed in Solar Orbiter. (Bottom) Back-mapped points from the high speed stream that was observed from Solar Orbiter on 13/05/2021 (blue stars), the background is the AIA 193 $\AA$ synoptic map.}
    \label{fig:solo_plot}
\end{figure}

To illustrate this we present in Fig.~\ref{fig:solo_plot} an example of the back-mapping achieved for an in situ high speed stream observed by Solar Orbiter on 13/05/2021. 
During this stream 23 in situ points were selected and back-mapped at their source location on the solar surface. These points can be seen as blue stars at the bottom panel of Fig.~\ref{fig:solo_plot}. All of them are connected to the edges of a small coronal hole and the fact that they overlap is due to the small separation between them, which is not visible in the full synoptic map representation. 
A more detailed analysis of these and future events will be performed in follow-up studies.

We must also note here that there is an underlying caveat in the derived uncertainty of the source location based on the velocity profile analysis. As \citet{Nolte1973a} indicated, there are two competing effects close to the Sun: solar wind acceleration (tends to move the back-mapped point Westwards; higher longitude) and corotation (tends to move the back-mapped point Eastwards; lower longitude) that approximately cancel out. This makes the use of a constant speed a valid approximation for back-mapping. Now, in our framework, the velocity profiles do not have an azimuthal component above the source surface, i.e. we consider acceleration without corotation. This actually implies that the solar wind acceleration is the dominant effect above the SS (below the SS the magnetic mapping, on the contrary, involves that the plasma fully corotates with the Sun). 
The SS  therefore consists in a sharp transition from a fully corotational solar wind to a solar wind with no azimuthal velocity component. But theoretical works \citep{Weber1967} and observational evidence from Parker Solar Probe indicate that there still exists a residual azimuthal velocity component at heights above the SS. Moreover, this azimuthal component is expected to increase as a function of radial distance (close to the Sun) and reach its maximum value below the Alfv\'en surface \citep{Weber1967}, which is typically well above the height of the source surface (i.e. 10-40\solrad{}).  
On the other hand, in situ measurements close to 1 AU show very small values of $v_{\phi}$ for the fast solar wind \citep[in the order of a couple \kms, and even negative values;][]{Lazarus1971,Pizzo1983}. Furthermore, modeling analysis shows that absolute $v_{\phi}$ values of the solar wind are an order of magnitude smaller than $v_r$ values from about 2-3 \solrad{} until 1 AU \citep{Keppens1999}. Although it does not reflect directly the effect on the back-mapped locations (the integration of the velocity profile is necessary for that), this comparison of absolute values of the two velocity components is a first indication of the contribution that the two components have in the propagation of the solar wind. 
Less crude approximations (for example using MHD models of the solar corona) would take this component into account. But if we want to remain in a back-mapping framework that is not computationally heavy and has short run times, other avenues for investigating the effect of the azimuthal component must be explored. 

We therefore first tried to account for the effect of corotation on the back-mapped points (above the SS) by using a similar methodology as for the radial velocity profile space (see Sect.~\ref{velocity_profiles}), where we parameterize the velocity profiles and then constrain them by real observations. For this we followed the work of \citet{MacNeil2022} and derived azimuthal velocity profiles based on the model of \citet{Weber1967} (see Eq.~\ref{eq-Weber_Davis} in the Appendix) and a range of locations for the Alfv\'en point (see Appendix \ref{v_phi_appendix} for more details). Then we compared these azimuthal velocity profiles with real measurements of $v_{\phi}$. 
Measurements of the azimuthal component of the solar wind velocity close to the Sun were only made possible recently with Parker Solar Probe. But these first measurements have shown surprising results with extremely large $v_{\phi}$ values, almost an order of magnitude larger than what was theoretically expected \citep{Kasper2019a}. 
The scarcity of $v_{\phi}$ measurements closer than 1 AU, together with their large uncertainty, made it impossible for us to use real observations to constrain the azimuthal velocity profile space. Comparison of the $v_{\phi}$ profiles with the observations can be seen in Fig.~\ref{fig:appendix}.

\begin{figure}[h]
    \centering
    \includegraphics[]{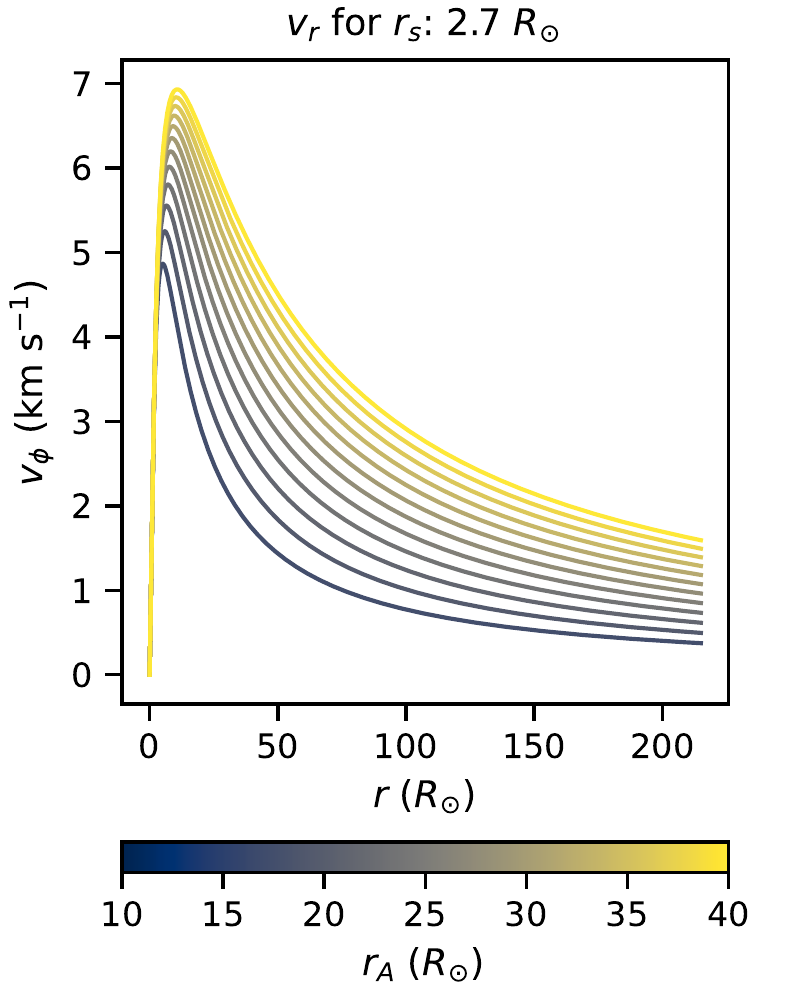}
    \caption{Family of azimuthal velocity profiles for the solar wind, derived using the hybrid radial velocity profile with $r_s = 2.7$, for a range of Alfv\'en point locations as indicated in the colorbar at the bottom. For these profiles an in-situ solar wind speed of 600 \kms at L1 ($\sim$ 215 \solrad{}) was considered.}
    \label{fig:vphi_plot}
\end{figure}

Consequently, in order to estimate the effect of corotation on the back-mapped points at the source surface we have to content ourselves with only using the location of the Alfv\'en point as a free parameter, taken inside a reasonable range.   
We select the range from 10 to 40 \solrad{}, as we try to encompass most of the values that are present in literature \citep{Weber1967,Keppens1999,Riley1999,Zhao2010,DeForest2014,DeForest2016,Kasper2019a,Kasper2021}.
Now we can compute a family of azimuthal profiles for a given $v_r$ profile.
For this exemplification we consider the radial velocity profile with $r_s = 2.7$ \solrad{}, from the example of hybrid radial velocity profiles presented in Sect.~\ref{Ballistic_Mapping} ($r_{s/c} = 215$ \solrad{} (L1), $v_{insitu} = 600$ \kms). This profile is selected as it is one of the $v_r$ profiles that produces azimuthal profiles with large $v_{\phi}$ values, and it is still compatible with solar wind observations close to the Sun. 
The resulting family of azimuthal profiles is shown in Fig.~\ref{fig:vphi_plot}.

Using these azimuthal profiles we can calculate the longitudinal displacement at the source surface $\Delta\Phi_{ss}^{s/c}$ (by evaluating the integral of Eq.~\ref{eq:delta_phi_integral}, see Appendix), i.e. the change in heliographic Carrington longitude of a solar wind parcel from the point of its release (SS) and the point that is measured in situ (S/C).
These profiles result in a $\Delta\Phi_{ss}^{s/c}$ in the range 41.6-42.7$^{\circ}$ (depending on the location of the Alfv\'en point), compared to a $\Delta\Phi_{ss}^{s/c}$ of 48.7$^{\circ}$ if only the radial velocity profile is considered. This shows that back-mapped footpoints on the source surface can be moved up to 6-7$^{\circ}$ Eastward, when the effect of corotation is taken into consideration, between the source surface and the spacecraft (assuming an azimuthal velocity component from the model of \citet{Weber1967}). It is worthwhile to notice that this represents the potential maximum effect of corotation in this model, because we consider Alfv\'en point locations as far as 40~\solrad{} and because we used the radial velocity profile that produces the most Eastward displacement of the back-mapped points.

Let us now try to examine what would be the effect in the uncertainty region on the solar surface when we back-map considering an accelerating solar wind velocity profile that has both $v_r(r)$ and $v_\phi(r)$ components.
To accomplish that we need to look first at the longitudinal difference at the source surface ($\Delta\phi_{ss}$) between a back-mapped point with 1) a constant speed profile ($v_r(r)=const,\ v_\phi(r)=0$), 2) an accelerating velocity profile but no azimuthal component ($v_r(r)\neq const,\ v_\phi(r)=0$) and 3) an accelerating velocity profile with both radial and azimuthal components ($v_r(r)\neq const,\ v_\phi(r) \neq 0$).
To compute this difference we examine the three events that we studied and for every in situ point in these events we compute the location at the source surface of a back-mapped point for which we considered 1) a wind profile with constant radial speed equal to the in situ measured speed, 2) a wind profile with a radial speed derived from one of our hybrid profiles by selecting the maximum $r_s$ value that still produces a profile compatible with remote sensing observations ($v_r^{max\ r_s}(r)$) and 3) a wind profile with the same radial speed ($v_r^{max\ r_s}(r)$) but also an azimuthal component $v_\phi(r)$ that was computed from this specific radial profile using the model of \citet{Weber1967}.
With these locations we can calculate the average longitudinal difference at the source surface, $\Delta\phi_{ss}$, between the ballistically back-mapped points (case 1 from above) and the back-mapped points with an accelerating radial profile (case 2 from above; largest $r_s$). The result is a $\Delta\phi_{ss} = + 9.02^{\circ} \pm 0.31^{\circ}$. Note here the difference between the $\Delta\phi_{ss}$ and the $\Delta\Phi_{ss}^{s/c}$ mentioned earlier, with the former representing the difference in longitude at the source surface and the later the difference in longitude between the source surface and the location of the spacecraft that made the in situ measurement.   
Following the same methodology, we can compute the average longitudinal difference at the source surface between the ballistically back-mapped points (case 1 from above) and the back-mapped points with velocity profile that has both radial and azimuthal component (case 3 from above; largest $r_s$ and $v_\phi(r)$). 
The result is $\Delta\phi_{ss} = + 3.06^{\circ} \pm 0.16^{\circ}$. 
We must clarify here that for the derivation of the azimuthal profiles we considered Alfv\'en point locations ($r_A$) inside our whole range (10-40~\solrad{}), but for the calculation of the average we took only the $v_\phi(r)$ profile with the $r_A$ that produced the biggest displacement from the ballistic footpoint.
Additionally, the positive sign in the $\Delta\phi_{ss}$ values represents that the longitudinal difference will be Westward and the negative sign that it will be Eastward (lower longitude values).


These results indicate that the back-mapped location at the source surface for an accelerating velocity profile ($v_r(r) \neq const,\ v_\phi(r) = 0$) can never be Eastward of the ballistically back-mapped location ($v_r(r) = const$). From our analysis this is also true if we include an azimuthal component which is derived from the $v_r(r)$ profile and a certain Alfv\'en point location (\citet{Weber1967} model). At a first glace, this result might seem in contradiction to the $\pm 10^{\circ}$ uncertainty in the projected longitude of the ballistic mapping that \citet{Nolte1973a} derived. But \citet{Nolte1973a} computed this uncertainty by comparing the Carrington longitude of the ballistic footpoint at $r=0$ with the Carrington longitude resulting from an accelerating velocity profile down to $r_0$ (the release zone of the wind), considering rigid body corotation with the Sun below this point (i.e an equivalent of the SS in our framework). The values they selected for $r_0$ were 21.5 \solrad{} (0.1 AU) and 53.7 \solrad{} (0.25 AU), which can be considered quite large for the transition of the solar wind to a purely radial propagation. Furthermore, this uncertainty ($\pm 10^{\circ}$), which is typically cited in solar wind back-mapping applications, is appropriate only when the ballistic mapping is considered until the solar surface but when there is a two step back-mapping (ballistic+magnetic) is not applicable anymore. In the latter the source surface height plays an important role as it indicates where the corotation with Sun stops. Furthermore, the uncertainty of the back-mapped locations at the source surface is better derived by the uncertainty in the velocity profile that was used, and the uncertainty at the solar surface after tracing the magnetic field lines (from source surface to solar surface).

In that respect, we have shown that the uncertainty of the back-mapped position at the source surface, due to the velocity profile of the fast solar wind, can be 0-9$^{\circ}$ Westward of the ballistic point when using an accelerating velocity profile (no azimuthal component) and 0-3$^{\circ}$ when also including an azimuthal component. 
Consequently, this reduction in the longitudinal spread of the back-mapped points at the source surface ($\Delta\phi_{ss}$), due to $v_{\phi}$, results in a reduction of the uncertainty area at the solar surface ($\le 1$ deg) compared to the case where we consider only $v_r$.
A schematic example of this reduction in the uncertainty area at the solar surface when we consider corotation effects above the source surface can be seen in Fig.~\ref{fig:appendix_cartoon}

Despite the fact that this analysis of the azimuthal component in the solar wind is a first approximation, it provides a rough estimate of the maximum effect that corotation can have on the back-mapped points close to the Sun, as explained above.
More in depth analysis of the azimuthal velocity component in the solar wind ($v_{\phi}$) and its effect on the back-mapping would require additional observational signatures or more advanced models, like MHD. Such an investigation is outside the scope of the current work, since the main effect of including an azimuthal component is a reduction in the uncertainty area due the choice of velocity profile (compared to having only a radial component). Also, this does not affect the hierarchical ordering in the significance of each source of uncertainty that we analyzed so far.

\section{Conclusions}\label{Conclusions}
We present a framework for estimating the uncertainty in the back-mapping of the fast solar wind, focusing on 3 main sources of uncertainty.
These are the uncertainty in the velocity profile of the solar wind for the ballistic mapping step of the back-mapping, the height of the source surface in the PFSS model and the noise in the input synoptic magnetogram that is used for the magnetic field extrapolation, in the magnetic step of the back-mapping.
For the uncertainty in the velocity profiles (ballistic step), we have computed custom profiles which can be used to sample the velocity profile space.
Next, we derive the boundaries in the velocity profile space, on one hand from the constant speed propagation and on the other from an analytical profile constrained by Doppler dimmings observations close to the Sun. A sampling in the velocity profile space translates into different propagation times, meaning different locations on the source surface and subsequently an uncertainty in the source region on the solar surface. 
For the height of the source surface a range of acceptable heights is taken and the back-mapping, of the same in situ point, is computed for all of them. The radial distance of the source surface has a significant impact in the underlying magnetic field topology, resulting in a uncertainty in the back-mapped position.
Lastly, we examined the effect of noise in the synoptic magnetogram, by back-mapping a given in situ point for different realizations of noise in the input magnetogram. The magnetogram noise has an additional impact in the magnetic connectivity from the source surface down to the solar surface, creating an additional uncertainty in the source region of the fast solar wind.

To group the back-mapped points from all sources of uncertainty, a clustering method called Gaussian Mixture Model was used. This gives us the ability to calculate independent confidence ellipsoids for each cluster, essentially providing an uncertainty area for the source location of the fast solar wind. From this, additional metrics can be derived, such as the number of clusters, the distance of cluster centroids to the barycenter and the area of the most probable cluster, which are used to compare the back-mapping across multiple in situ points.

For this work we focused mainly on a single high speed stream, which was used as a case study to present our framework. From this we selected a single in situ point, analyzed the effect of all the sources of uncertainty and investigated the results, both visually by correlation with the EUV images and statistically with the quantities computed from the GMM. The results showed that the height of the source surface produces the biggest uncertainty in the final back-mapped position, followed by the sampling of the velocity profile space and the noise in the input magnetogram.
Subsequently, we built up statistical significance of these results by first analyzing all in situ points of our case study high speed stream and then performing the same analysis on two additional high speed streams. 
The hierarchy in the impact of each uncertainty source, that was found for a single in situ point persisted through this multi-event, statistical analysis. We also showed how modern data from Solar Orbiter can successfully be used in our framework.

To our knowledge such a detailed inspection of the uncertainties in the back-mapping of the fast solar wind has never been performed before, despite the fact this is one of the most used processes in the study of the solar wind. 
The main impact of this study lies in the precise estimation of each uncertainty source impact on the source region of the solar wind and the ability of this framework to produce a confidence area around the back-mapped position at the solar surface. These are especially useful in the context of linkage analysis where remote sensing observations are combined with in situ measurements for the study of solar wind. For example if we have some spectral observations at the time of back-mapping, the overlapping of the uncertainty ellipsoids with the spectral field of view can indicate which part of the observations is more suitable in the context of connection to in situ parameters. Lastly, this framework can be extended to the study of slow solar wind too. Future improvements may use actual MHD coronal models to replace the PFSS-based magnetic mapping. The framework is implemented in python and is considered to be made available to the community.

\begin{acknowledgements}
    A. Koukras is supported by the Research Foundation - Flanders (FWO), PhD-aspirant grant (1180919N).
    L. Dolla thanks the Belgian Federal Science Policy Office (BELSPO) for the provision of financial support in the framework of the PRODEX Programme of the European Space Agency (ESA) under contract no. 4000136424.
    R. Keppens acknowledges the support by the European Research Council (ERC) under the European Unions Horizon 2020 research and innovation program (grant agreement No. 833251 PROMINENT ERC-ADG 2018) and by Internal KU Leuven funds, through the project C14/19/089 TRACESpace. 
    This work uses data from the National Solar Observatory Integrated Synoptic Program, which is operated by the Association of Universities for Research in Astronomy, under a cooperative agreement with the National Science Foundation and with additional financial support from the National Oceanic and Atmospheric Administration, the National Aeronautics and Space Administration, and the United States Air Force. The GONG network of instruments is hosted by the Big Bear Solar Observatory, High Altitude Observatory, Learmonth Solar Observatory, Udaipur Solar Observatory, Instituto de Astrof\'{\i}sica de Canarias, and Cerro Tololo Interamerican Observatory.
    This research has made use of SunPy, an open-source and free community-developed solar data analysis package written in PYTHON \citep{Mumford2015}. This research made use of HelioPy, a community-developed PYTHON package for space physics \citep{Stansby2020heliopy}. The synoptic AIA maps were created using the open source code \emph{solarsynoptic}\footnote{https://solarsynoptic.readthedocs.io/en/latest/}, developed by David Stansby.
    All figures were produced using the MATPLOTLIB plotting library for PYTHON \citep{Hunter2007}.

\end{acknowledgements}

\bibliography{Solar_Wind.bib}  

\begin{appendix}\label{Appendix}
    \section{Azimuthal velocity analysis}\label{v_phi_appendix}
        As indicated in Sect.~\ref{Ballistic_Mapping} and in Sect.~\ref{Discussion} the effect of corotation of the solar wind close to the Sun is represented by a non zero azimuthal velocity component $v_{\phi}$ that would affect the derived longitudinal displacement on the source surface after back-mapping.
        In order to examine this effect in more detail we compute a family of parameterized azimuthal velocity profiles, following the methodology indicated in \citet{MacNeil2022}. In this paper the expression for $v_{\phi}(r)$, from \citet{Weber1967}, is used based on pre-calculated $v_r(r)$ radial velocity profiles:
        
        \begin{equation}
            v_{\phi}(r) = \frac{\Omega r}{v_A} \frac{v_A - v_r(r)}{1 - M_A^2(r)}, \label{eq-Weber_Davis}
        \end{equation}
        where $v_A$ is the local Alfv\'en speed ($v_A = v_r(r_A)$;with $r_A$ the location of the Alfv\'en point) and $M_A(r)$ the radial Alfv\'en Mach number, with $M_A^2(r) = (v_r r^2)/(v_A r_A^2)$.      

\begin{figure*}[h]
            \centering
            \includegraphics[scale=0.41]{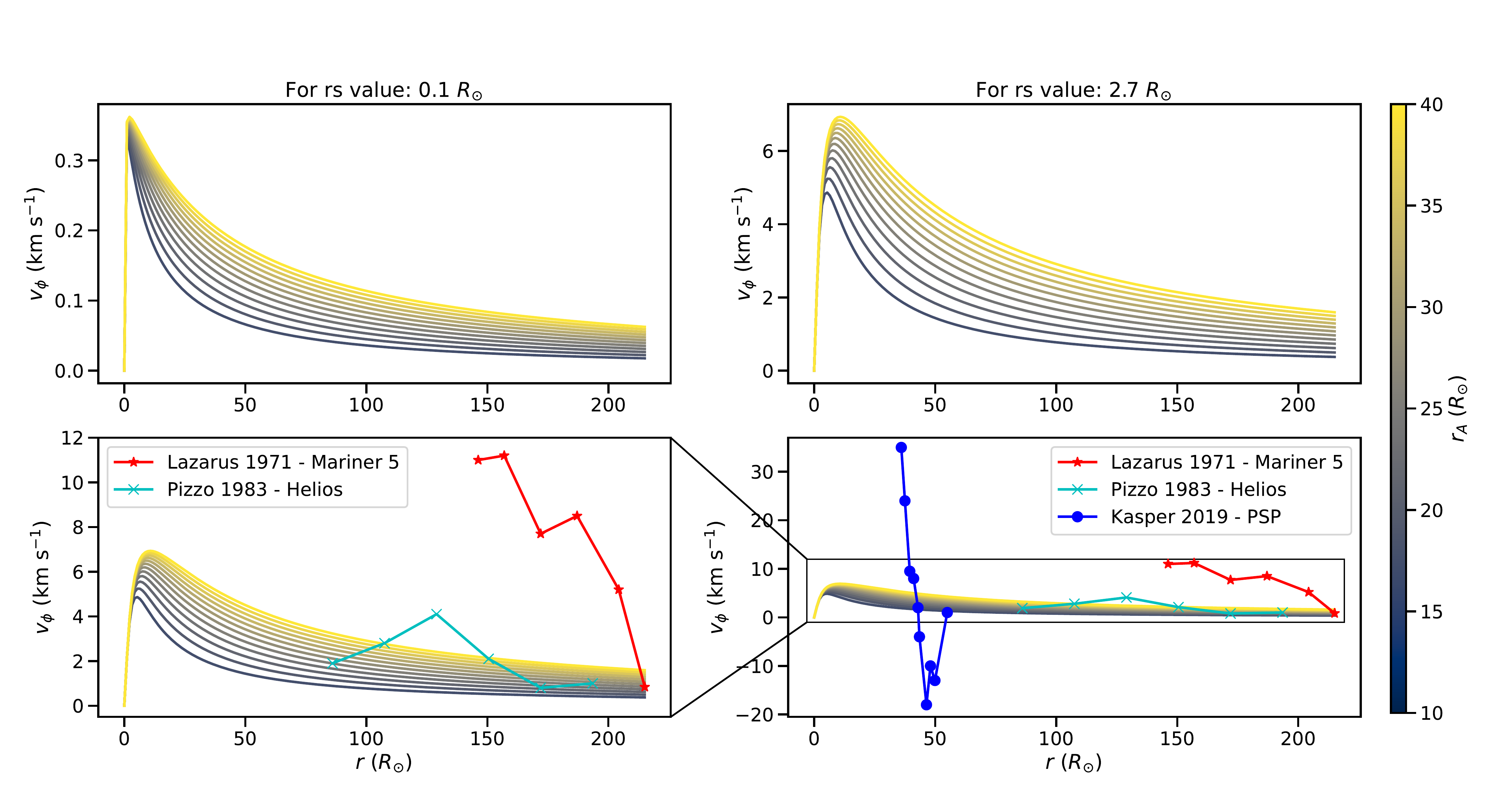}
            \caption{Family of azimuthal velocity profiles for the solar wind, derived using different hybrid velocity profiles ($v_r$) and for a range of Alfv\'en point locations as indicated by the colorbar on the right. (Top left) Azimuthal velocity profiles using a hybrid velocity profile with $r_s = 0.1$. (Top right) Azimuthal velocity profiles using a hybrid velocity profile with $r_s = 2.7$. (Bottom left) The same family of profiles as the top right panel but including earlier measurements of $v_{\phi}$ as indicated in the legend. (Bottom right) The same as bottom left but including newer measurements of $v_{\phi}$ from Parker Solar Probe. The black rectangle indicates the field of view of the plot on the left.}
            \label{fig:appendix}
        \end{figure*}

        In this formulation the two parameters that control the shape of the azimuthal velocity profile are the location of the Alfv\'en point and the profile $v_r(r)$. For the latter we can use the hybrid velocity profiles described in Sect.~\ref{velocity_profiles}. As for the location of the Alfv\'en point there has been a lot of discussion about its position, with more recent works indicating that it lies lower (smaller radial distances) than argued in the past \citep{DeForest2014,DeForest2016,Kasper2019a,Kasper2021,Bandyopadhyay2022}.
        A reasonable approach in this case would be to use a range of Alf\'en point locations that encompasses most the values mentioned in literature. An example of a family of azimuthal velocity profiles based on a range of Alfv\'en points between 10-40 \solrad{} and two different hybrid velocity profiles, with $r_s$ = 0.1 and 2.7 can be seen in top right and top left panels, respectively, of Fig.~\ref{fig:appendix}. For these profiles an in-situ solar wind speed of 600 \kms at L1 was used, similar to the example presented in Sect.~\ref{velocity_profiles}.
        
        The next step is to try to constrain this profile space by real measurements of $v_{\phi}$, similarly to the methodology that we used for the hybrid radial velocity profiles. Unfortunately $v_{\phi}$ measurements in distances smaller than 1 AU are very rare and often are accompanied by large uncertainties. The first $v_{\phi}$ measurements showed a fairly good agreement with the \citet{Weber1967} model for distances $\sim$ 0.8-1 AU, as reported from \citet{Lazarus1971} with Mariner 5 observations and \citet{Pizzo1983} with Helios observations. 
        But more recent measurements with Parker Solar Probe, which reached even closer to the Sun, showed surprisingly large $v_{\phi}$ values that are almost an order of magnitude higher than what would be expected from \cite{Weber1967} model. For the first two PSP encounters, \citet{Kasper2019a} reported values of up to $\sim{}$40 \kms when PSP was located around 36 \solrad. 
        These measurements are plotted on top of our velocity profiles as seen in the bottom panels of Fig.~\ref{fig:appendix}. The bottom left panel displays the azimuthal velocity profiles calculated with $r_s = 2.7$ (this $v_r$ provides the largest $v_{\phi}$ values) together with the earlier observations of the azimuthal velocity and the bottom right panel the same family of profiles but including the newest observations from PSP, as indicated in the legend. The black rectangle in the bottom right panel indicates the region that is zoomed-in in the bottom left panel. 
        
        It becomes clear that these observations can not be used to constrain the azimuthal velocity profile space, since to have a profile that matches the PSP observations, extremely large distances for the location of the Alfv\'en point are needed. Additionally, there is not a clear consensus about the uncertainty of these new $v_{\phi}$ measurements and if these large values should be trusted or there are underlying effects and corrections that should be taken into consideration (for a more detail discussion see \citet{MacNeil2022} and references therein). This hinders the use of this methodology to constrain the azimuthal profile space and derive a measure of the effect of corotation in back-mapped locations, at least until more $v_{\phi}$ observations become available and their robustness increase. 
        
        Despite the above, a rough estimate about the effect of corotation close to the Sun can be achieved if we constrain the azimuthal velocity profile space only by the range of Alfv\'en points locations. In this approach, the $v_r(r)$ will be selected as the radial velocity profile that agrees with observations (Doppler dimming) and produces $v_{\phi}(r)$ profiles with the largest values, making the only free parameter the location of the Alfv\'en point. 
        Consequently, for a range of Alfv\'en points we can compute the effect in the back-mapped location, indicated as the longitudinal displacement in Carrington coordinates $\Delta\Phi_{ss}^{s/c}$, using 
        
        \begin{equation}
            \Delta\Phi_{ss}^{s/c} = \int_{r_{ss}}^{r_{sc}} \frac{\Omega - \frac{v_{\phi}(r)}{r}}{v_r(r)}, \label{eq:delta_phi_integral}
        \end{equation}
        similar to Eq.1 of \citet{MacNeil2022}. With $r_{ss}$ and $r_{sc}$ being the heliocentric distance of the source surface and of the spacecraft that measures the in situ wind, respectively.
        
        As a conclusion, with this longitudinal displacement we are able to estimate an upper limit (due to the specific selection of $v_r(r)$) for the effect of corotation close to the Sun, under the consideration of three physically informed assumptions: the \citet{Weber1967} model, a $v_r(r)$ that correlates with remote sensing observations and a reasonable range of locations for the Alfv\'en point.
        
        \begin{figure*}[h]
            \centering
            \includegraphics[width=\textwidth]{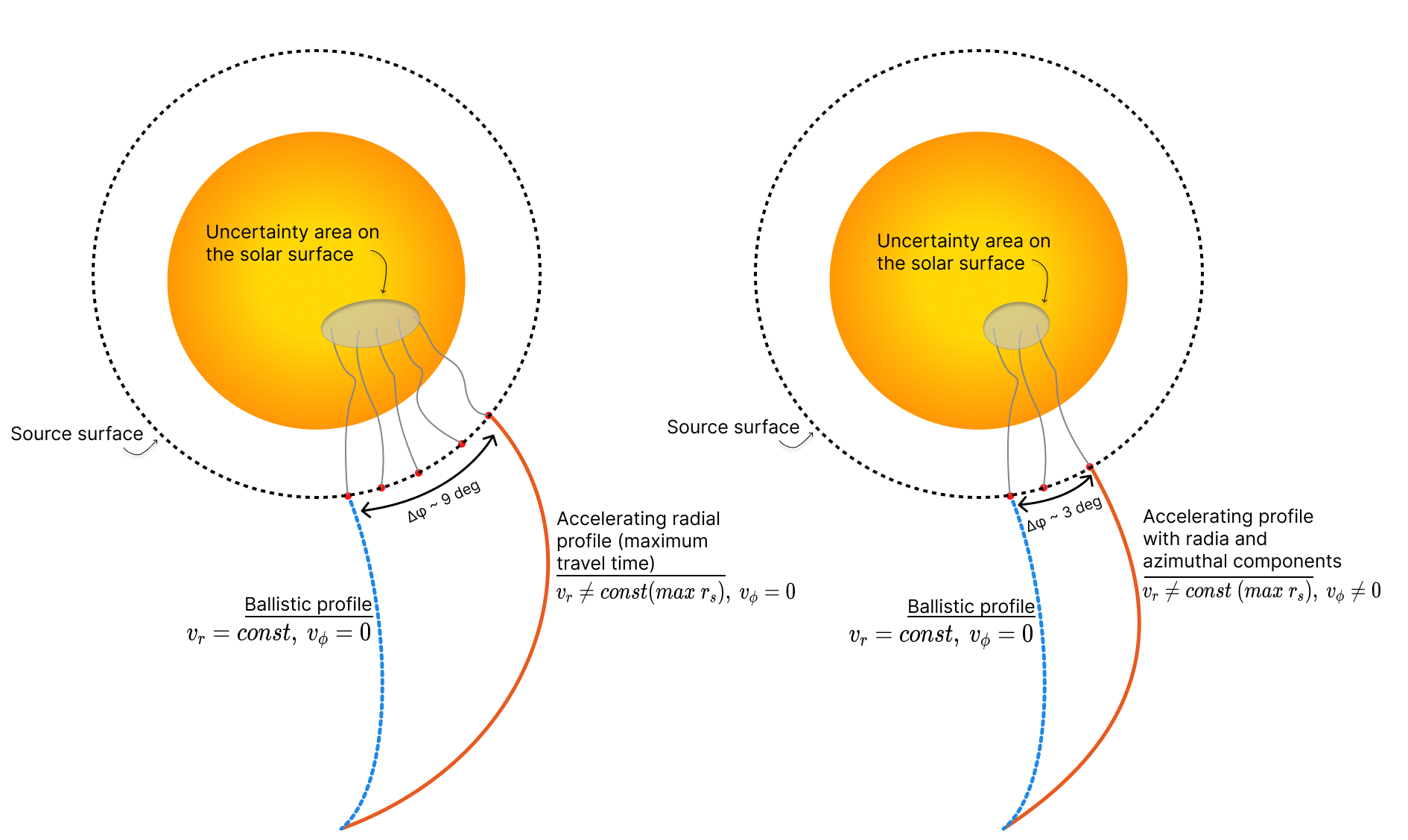}
            \caption{Cartoon representing the effect in the uncertainty area on the solar surface, if corotation effects above the source surface are taken into consideration. The longitudinal spread of back-mapped points at the source surface is determined by the location of the ballistically back-mapped point and that of an accelerating velocity profile with the largest travel time (i.e. a profile which produces the largest offset at the source surface). In both illustrations the location of the ballistically back-mapped point at the source surface is given by the dashed blue line. In the illustration on the left, the back-mapped location at the source surface with the largest longitudinal offset is given by an accelerating velocity profile, derived from the family of hybrid profiles presented in Sect.~\ref{velocity_profiles} and for the largest $r_s$ value that produces a velocity profile compatible with remote sensing observations. In the illustration on the right, the location at the source surface of the back-mapped point with the largest longitudinal offset is given by an accelerating profile that has both a radial and an azimuthal component. The radial component is the same as the schematic on the left and the azimuthal component is derived from the \citet{Weber1967} model as discussed in Sect.~\ref{Discussion} and in the Appendix.
            It is noteworthy that the longitudinal spread of back-mapped points at the source surface is reduced when an azimuthal component is added to the accelerating velocity profile. Furthermore, this reduced longitudinal spread is included in that computed when considering velocity profiles without an azimuthal component. Consequently, the uncertainty area on the solar surface, computed after tracing the field lines that connect the points back-mapped at the source surface to the solar surface, will also reduce. 
            This simple cartoon represents more closely the case of a coronal hole with super-radial expansion (similar to the main morphology of the events in our study), but the reduction in the uncertainty area on the solar surface should be in principle independent of the magnetic field topology. The reason for this is that the longitude locations (of the back-mapped points at the source surface) when we consider an azimuthal component are a subset of the ones when we consider only a radial component.}
            \label{fig:appendix_cartoon}
        \end{figure*}
\end{appendix}

\end{document}